%% file: draft_eta_eta_jpsi.tex
\documentclass[aps,prd,reprint,superscriptaddress,floatfix,showpacs]{revtex4-2}
\usepackage{graphicx}
\usepackage[colorlinks,urlcolor=blue, citecolor=blue,linkcolor=blue]{hyperref}
\usepackage[capitalise]{cleveref }
\usepackage[english]{babel}
\usepackage{placeins}
\usepackage{longtable}
\usepackage{dcolumn}
\usepackage{xcolor}
\usepackage{overpic}
\usepackage{siunitx-hep}
\usepackage{longtable}

\newcommand{\BESIIIorcid}[1]{\href{https://orcid.org/#1}{\hspace*{0.1em}\raisebox{-0.45ex}{\includegraphics[width=1em]{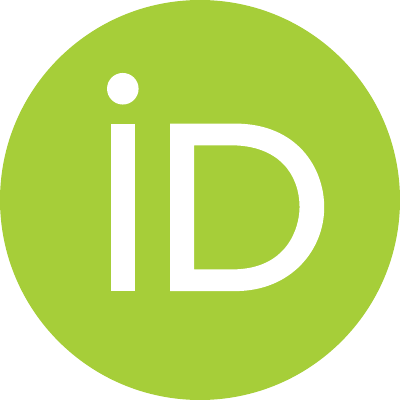}}}}

\newcommand{\elp}{e^{+}}
\newcommand{\elm}{e^{-}}

\newcommand{\mup}{\mu^{+}}
\newcommand{\mum}{\mu^{-}}

\newcommand{\lp}{l^{+}}
\newcommand{\lm}{l^{-}}

\newcommand{\jpsi}{J/\psi}
\newcommand{\psip}{\psi(2S)}

\newcommand{\pip}{\pi^{+}}
\newcommand{\pim}{\pi^{-}}
\newcommand{\piz}{\pi^{0}}

\newcommand{\pb}{\bar{p}}

\newcommand{\br}{\mathcal{B}}

\newcommand{\chisq}{\chi^{2}}
\newcommand{\dbs}{\bar{D}^*}

\newcommand{\twogg}{\ensuremath{\eta \to \gamma\gamma}}
\newcommand{\threepi}{\ensuremath{\eta \to \pip\pim\piz}}

\graphicspath{{./}}

\begin{document}


\title{\boldmath Search for the reaction channel $\elp \elm \to \eta\eta\jpsi$ and the isospin partner of the $Z_c(3900)$ at center-of-mass energies $\sqrt{s} = 4.226-4.950$~GeV }

\include{Authorlist}

\begin{abstract}
We search for the reaction channel $\elp\elm\to\eta\eta\jpsi$ in a data sample with center-of-mass energies from 4.226 to 4.950~GeV that was collected by the BESIII detector operating at the Beijing Electron Positron Collider (BEPCII). The data analysis is performed with two different reconstruction methods, exclusive and semi-inclusive, enabling a comparison and combination of the results. Only at a few energy points a non-zero cross section is observed with a statistical significance of more than $3\sigma$ using one of the methods. Therefore, the corresponding upper limits of the cross section at the 90\% confidence level are determined. The energy dependent results show clear deviations from the the line shape expected from three-body phase space alone. Since the statistical significance for almost all center-of-mass energies is low, the upper limits for the reaction channel $\elp\elm\to\eta\eta\jpsi$ also serve as limits for the existence of a possible isospin 
partner to the charmonium-like isospin triplet $Z_{\rm c}(3900)$ which decays to  $\jpsi \eta$.
\end{abstract}

\date{\today}


\maketitle

\section{Introduction}
The nature of the charmonium-like states is a hot topic in hadron physics, since
those states do not fit into the spectrum of the conventional quark model~\cite{Barnes:2005pb} and
challenge the understanding of charmonium spectroscopy and QCD calculations~\cite{Brambilla:2010cs,Briceno:2015rlt}.
In the past two decades, a series of charmonium-like states have been discovered by several experiments. 
Among these, the states $Y(4260)$ and $Y(4360)$ do not fit in the charmonium spectrum above the $D\bar{D}$ threshold because all states are already covered by the well established charmonium states $\psi(3770)$,
$\psi(4040)$, $\psi(4160)$, and $\psi(4415)$~\cite{ParticleDataGroup:2024cfk}.
So far, no decay to $D\bar{D}$ of these $Y$-states has been observed, although
their masses are above $D\bar{D}$ threshold~\cite{BaBar:2006qlj,Belle:2007qxm}.

The $Y(4260)$ state was discovered in the reaction channel $\elp\elm\to\pip\pim\jpsi$
by the BaBar collaboration~\cite{BaBar:2005hhc} using the initial state radiation (ISR) process
and then confirmed by the CLEO~\cite{CLEO:2006ike} and Belle~\cite{Belle:2007dxy} collaborations.

Several theoretical interpretations have been proposed for the $Y(4260)$ such as
tetraquark~\cite{Maiani:2005pe}, meson molecule~\cite{Ding:2008gr, Wang:2013cya},
hadroquarkonium~\cite{Alberti:2016dru, Li:2013ssa},
hybrid meson~\cite{Zhu:2005hp, Close:2005iz, Kou:2005gt},
and others~\cite{Guo:2013sya, Maiani:2013nmn, Braaten:2013boa, Liu:2013vfa}. Based on the study of its fine structure by the BESIII collaboration via $\elp \elm \to \pip \pim \jpsi$~\cite{BESIII:2016bnd} with high statistics data,
$Y(4260)$ is now regarded as an overlap of two resonances: $Y(4220)$ and $Y(4320)$.

The invariant $\pip \pim \psip$ mass spectrum was studied by the BaBar collaboration,
where the $Y(4360)$ was observed~\cite{BaBar:2006ait}. The
Belle collaboration confirmed its existence~\cite{Belle:2007umv},
and the BESIII collaboration performed a study of its fine structure based on $\elp \elm \to \pip \pim \psip$~\cite{BESIII:2017tqk}.
Like $Y(4260)$, the fine structure study indicates that $Y(4360)$ is a combination of two resonances:
one is around the mass of 4.220 GeV/$c^{2}$, and another around the mass of 4.360 GeV/$c^{2}$~\cite{Zhang:2017nqg}.
In confirmation of the $Y(4360)$, the Belle collaboration observed another resonance $Y(4660)$ in the invariant $\pip \pim \psip$ mass spectrum~\cite{Belle:2007umv},
and this was confirmed by BaBar~\cite{BaBar:2012hpr} and BESIII~\cite{BESIII:2017tqk}.
This is considered as a single state and could be an excellent candidate for a $[cd][\bar{c}\bar{d}]$ diquark-antidiquark bound state~\cite{Cotugno:2009ys},
for an $f_{\rm 0}(980)\psip$ bound state~\cite{Guo:2008zg}, or an $f_{\rm 0}(980)\eta_{\rm c}^{\prime}$ bound state~\cite{Guo:2009id}.

Recently, with the large $\elp\elm$ annihilation data samples below 4.950~GeV, the BESIII collaboration observed the two states $Y(4500)$ and $Y(4710)$ in the reaction channel $\elp\elm\to K^+K^-\jpsi$~\cite{BESIII:2022joj,BESIII:kkjpsi}, while in the reaction channel  $\elp\elm\to K_S^0K_S^0\jpsi$ only evidence of the charmonium-like state at 4.710~GeV was found so far~\cite{BESIII:ksksjpsi}. More studies in the range of the $Y$ states, especially at large masses, are needed to confirm the structures where the first evidences are reported. 

Besides the $Y$ states, the heavy and charged $Z_c$ states have been observed. These properties make them good candidates for four-quark states, because of their decays to charmonia and their electric charge. The first confirmed state was the $Z_c(3900)^\pm$ decaying to $\jpsi\pi^\pm$, which was discovered by the BESIII collaboration in the reaction channel $\elp \elm \to \pip\pim\jpsi$ at $\sqrt{s} = 4.26$~GeV \cite{Zc3900BES}. 
This resonance was also observed by the Belle collaboration \cite{Belle:Z3900} and confirmed using CLEO-c data \cite{Xiao:Zc3900}. 
In the CLEO-c data set, first evidence for a neutral state $Z_c(3900)^{0}$ was found in $\elp \elm \to \piz\piz\jpsi$ \cite{Xiao:Zc3900}, which was later also observed by the BESIII collaboration~\cite{Ablikim:2015tbp}. 
Also, the decays of the neutral and the charged $Z_c(3900) \to D\dbs$ have been observed
\cite{Ablikim:2013xfr}, triggering the interpretation of this resonance to be either a tetra-quark state or a $D$-meson molecule \cite{Zc_mol}. In addition, the
$Z_c(4020)$ triplet was observed in the reaction channels $\elp \elm \to \pip\pim h_c$ and $\elp \elm \to \piz\piz h_c$, with $Z_c(4020) \to h_c\pi$~\cite{PhysRevLett.111.242001,Ablikim:2014dxl}. Furthermore, structures whose poles are compatible with $Z_c(4020) \to D^{*}\dbs$ were also found by the BESIII collaboration in the reaction channels $\elp \elm \to \pip(D^{*}\dbs)^{-}$ and $\elp \elm \to \piz(D^{*}\dbs)^{0}$~\cite{PhysRevLett.112.132001,PhysRevLett.115.182002}. 
The observation of the isospin triplet states $Z_c(3900)$ decaying to $\jpsi\pi$ and $Z_c(4020)$ decaying to $h_c\pi$ suggests that there might also exist isospin singlet states $Z'_c$ decaying to $\jpsi\eta$ or $h_c\eta$. Search for these singlets provides further information on the $Z_c$ states.  

In this analysis, we present a search for the reaction channel $\elp\elm\to\eta\eta\jpsi$ at center-of-mass (CM) energies from 4.226 to 4.950~GeV,
using the data samples with an overall time-integrated luminosity ($\mathcal{L}$) of 15.1~fb$^{-1}$~\cite{BESIII:2015qfd, BESIII:2020eyu, BESIII:2022xii, BESIII:2022ulv}
taken at twenty-nine energy values by the BESIII experiment operating at the BEPCII storage ring. This analysis is performed by exploiting two different methods: the exclusive method, where all final state particles are reconstructed, and the semi-inclusive method, where the $\jpsi$ and one $\eta$ are reconstructed and the second $\eta$ is treated as a missing particle.
Any observed events will also allow us to study possible resonant substructures in the $\jpsi\eta$ subsystem. 

\section{BESIII detector and data samples}
The BESIII detector~\cite{BESIII:2009fln} records symmetric $\elp\elm$ collisions provided by the BEPCII storage ring~\cite{Yu:2016cof}
in the CM energy range from 1.84 to 4.95~GeV, with a peak luminosity of $1.1 \times 10^{33}\;\text{cm}^{-2}\text{s}^{-1}$ achieved at $\sqrt{s} = 3.773\;\text{GeV}$.
The BESIII collaboration has collected large data samples in this energy region~\cite{BESIII:2020nme}. The cylindrical core of the BESIII detector covers 93\% of
the full solid angle and consists of a helium-based multilayer drift chamber (MDC),
a plastic scintillator time-of-flight system (TOF), and a CsI(Tl) electromagnetic calorimeter (EMC),
which are all enclosed in a superconducting solenoidal magnet providing a 1.0 T magnetic field.
The solenoid is supported by an octagonal flux-return yoke with resistive plate counter muon (MUC) identification modules interleaved with steel.
The charged-particle momentum resolution at 1 GeV/$c$ is 0.5\%, and the ${\rm d}E/{\rm d}x$ resolution is 6\% for electrons from Bhabha scattering.
The EMC measures photon energies with a resolution of 2.5\% (5\%) at 1 GeV in the barrel (end cap) regions.
The time resolution in the TOF barrel region is 68 ps, while that in the end cap region is 110 ps.
The end cap TOF system was upgraded in 2015 using multigap resistive plate chamber technology, providing a time resolution of 60 ps,
which benefits 71\% of the data used in this analysis~\cite{Cao:2020ibk}.

The search for the reaction channel $\elp \elm \to \eta \eta \jpsi$ is performed at 29 different CM energies between $4.22$ and $4.95$~GeV. The time-integrated luminosity
of these data sets ranges from $100$ to $1700$ pb$^{-1}$, with data collected over the past 10 years. For details, see Table \ref{tab:xsec}.

For the determination of the reconstruction efficiency, the background contribution, and the investigation of systematic uncertainties, several data samples were simulated with a {\sc geant4}-based~\cite{GEANT4:2002zbu} Monte Carlo (MC) toolkit, which includes the geometric description of the BESIII detector and its response. 
The simulation models the beam
energy spread and initial state radiation (ISR) in the $e^+e^-$
annihilations with the generator {\sc kkmc}~\cite{Jadach:1999vf, Jadach:2000ir}
All known particle decays are modeled with {\sc evtgen}~\cite{Lange:2001uf, Ping:2008zz} using
branching fractions taken from the Particle Data Group (PDG)~\cite{ParticleDataGroup:2024cfk}. Unknown decays are estimated with {\sc lundcharm}~\cite{Chen:2000tv, Yang:2014vra}.
Final state radiation from charged final state particles is incorporated using {\sc photos}~\cite{Richter-Was:1992hxq}.
MC simulated data samples (background MC), e.g. containing
light quark, open charm and charmonium final states, as well as $\elp\elm$ or $\mup\mum$ processes, are used to optimize the background suppression.
The reaction channel $\elp \elm \to \eta \eta \jpsi$ is used to optimize the background suppression and
is generated according to a phase-space distribution.
The {\sc kkmc} generator~\cite{Jadach:1999vf, Jadach:2000ir} is also used to determine the ISR correction factors, needed to convert the observed cross section
into the Born cross section~\cite{Ping:2013jka, Sun:2020ehv}.
In addition, several dedicated background reaction channels are simulated to study the influence and the best way to suppress the background.

\section{Introduction to data analysis}
The analysis of the reaction channel $\elp \elm \to \eta \eta \jpsi$ is performed in two different ways. The exclusive method is described in section~\ref{ref:ex}, while the semi-inclusive method is described in section~\ref{ref:in}. 

In both cases, the $\jpsi$ is reconstructed in both leptonic decays, $\jpsi \to \elp\elm$ (hereafter $ee$-mode) and $\jpsi \to \mup\mum$ (hereafter $\mu\mu$-mode).
For the exclusive method, both $\eta$ mesons are reconstructed exclusively, where the two decays $\twogg$ and $\threepi$ are used.  
For the semi-inclusive method, one $\eta$ has to decay to $\gamma\gamma$ while the other one is allowed to decay to anything. 
While the exclusive case provides almost background-free results, the semi-inclusive reconstruction method has higher detection efficiency.

Charged tracks detected in the MDC must satisfy a distance of closest approach to the interaction point (IP) less than $10$~cm along the $z$-axis ($|V_{\rm z}| < 10.0$~cm)
and less than $1$~cm in the transverse plane ($|V_{\rm xy}| < 1$~cm), and be reconstructed in the polar angle ($\theta$) range of $|\!\cos\theta| < 0.93$,
where the $z$-axis is the symmetry axis of the MDC, and $\theta$ is defined with respect to the $z$-axis.

Photon candidates are identified using showers in the EMC.
The deposited energy of the shower must be more than 25~MeV in the barrel region ($|\!\cos\theta| < 0.80$), and more than 50~MeV in the end cap region ($0.86 < |\!\cos\theta| < 0.92$).
To exclude  showers that originate from charged tracks,
the angle subtended by the EMC shower and the position of the closest charged track at the EMC must be greater than 10 degrees as measured from the IP.
To suppress electronic noise and showers unrelated to the event,
the difference between the EMC time and the event start time is required to be within $[0,\,700]$~ns.

\section{Exclusive method}
\label{ref:ex}
In the exclusive method, both $\eta$ mesons are reconstructed exclusively from the two decays $\twogg$ and $\threepi$.  

\subsection{Data analysis}
In addition to the basic track reconstruction criteria, the momentum of the lepton candidates is required to be greater than $1$~GeV/c, while for the pions it has to be lower than $1$~GeV/c.
The electrons and muons can be distinguished by using the quotient of energy deposit in the EMC divided by the measured momentum in the MDC; for electrons it has to be greater than 0.7$c$ and for muons lower than 0.3$c$. Since a photon pair is needed for the reconstruction of both $\eta$s, the number of photons in the event is required to be greater than three.

For the selection of $\piz$ and $\eta$ candidates decaying into two $\gamma$s, the combinations of two photon candidates whose
invariant mass is close to the meson nominal mass are retained (80 MeV/$c^{2} \le M(\gamma\gamma) \le$ 180 MeV/$c^{2}$ and 400 MeV/$c^{2} \le  M(\gamma\gamma) \le$ 700 MeV/$c^{2}$ for $\piz$ and $\eta$ candidates, respectively). Afterwards, a kinematic fit with $\piz$ or $\eta$ mass constraint is performed and a very loose $\chi^2$ cut is used to select the $\piz$ or $\eta$ candidates.
For the $\eta$ candidates which decay into $\threepi$, the invariant mass of the three pions must be inside the interval 400 MeV/$c^{2} \le M(\pip\pim\piz) \le$ 700 MeV/$c^{2}$.
In addition, a vertex fit is performed to the two charged pions, followed by a kinematic fit with constraints on the $\piz$ and $\eta$ masses. A very loose $\chi^2$ cut is used to select the $\eta$ candidates.

Every $\eta\eta\ell^{+}\ell^{-}$ combination is further analyzed after these pre-selections.
A vertex fit using all charged tracks is performed to ensure a common vertex.
In addition, a kinematic fit with constraints on energy and momentum conservation and on the masses of the two $\eta$s and of the $\jpsi$ is performed and a loose $\chi^2$ cut is used to select the candidates.
If more than one combination fulfils all selection criteria, the combination with the smallest $\chi^{2}_{\rm 7C}$ is selected.

\subsection{Background contribution}
The investigation of the background MC samples indicates that the following reaction channels contribute most to the remaining background events:

\begin{itemize}
\item{$\elp \elm \to K_{\rm S}^{0} K_{\rm S}^{0} \jpsi$ with
       $K_{\rm S}^{0} \to \piz\piz$, $K_{\rm S}^{0} \to \piz\piz$ 
       or with
       $K_{\rm S}^{0} \to \piz\piz$, $K_{\rm S}^{0} \to \pip\pim$};
\item{$\elp \elm \to \omega \chi_{\rm c1,2}$  with
      $\omega \to \pip \pim \piz$, $\chi_{\rm c1,2} \to \gamma \jpsi$
      or with
      $\omega \to \gamma \piz$, $\chi_{\rm c1,2} \to \gamma \jpsi$};
\item{$\elp \elm \to \piz \piz \jpsi$};
\item{$\elp \elm \to \piz \piz \psi(2S)$ with
      $\psi(2S) \to \jpsi \eta$
      or with
      $\psi(2S) \to \gamma \chi_{\rm c1,2}$, $\chi_{\rm c1,2} \to \gamma \jpsi$}.
\end{itemize}

The background channel $\elp \elm \to \piz\piz\jpsi$ is contributing most for the case where both $\eta$s decay to two photons. Here, the four photons from the $\piz$ decays are falsely assigned to the two $\eta$ candidates.
To suppress this contribution, we veto the combinations where the four selected photons are also able to form two $\piz$ candidates.
Then, the sum of the $\chi^{2}$ values of the kinematic fits from the two $\piz$ selections and the two $\eta$ selections is compared, and the hypothesis with the smaller sum is taken as a signal event.
This criterion suppresses \SI{93}{\percent} of the $\elp \elm \to \piz\piz\jpsi$ events while only \SI{0.2}{\percent} of the signal events are rejected.

The estimation of the background contribution of these channels is performed via MC samples. Some of the cross sections of all five background channels, necessary for the estimation, have been measured by the BESIII collaboration
\cite{ref:kkjpsi,ref:omegachic-421-442,ref:omegachic-4600,ref:pi0pi0jpsi-4600,ref:pi0pi0psi2s}.
Missing data points were interpolated in accordance with the fit functions provided in the publications. For each channel and each CM energy separately, large-size MC samples are produced, which correspond to multiples of the measured luminosity. They are used to determine the reconstruction efficiency of the events from each background channel, separately, with low statistical uncertainty. Having the cross section and the reconstruction efficiency, the number of expected background events for each channel is determined and scaled to the measured time-integrated luminosity. 
The results are shown in Table \ref{tab:xsec}. The unscaled number of all background events and the corresponding scaling factor (not listed), are used as input for the calculation of the upper limit in section \ref{sec:upperlimit}.

\subsection{Observed and Born cross section}
\begin{figure}[htbp]
  \centering
  \includegraphics[width=0.42\textwidth]{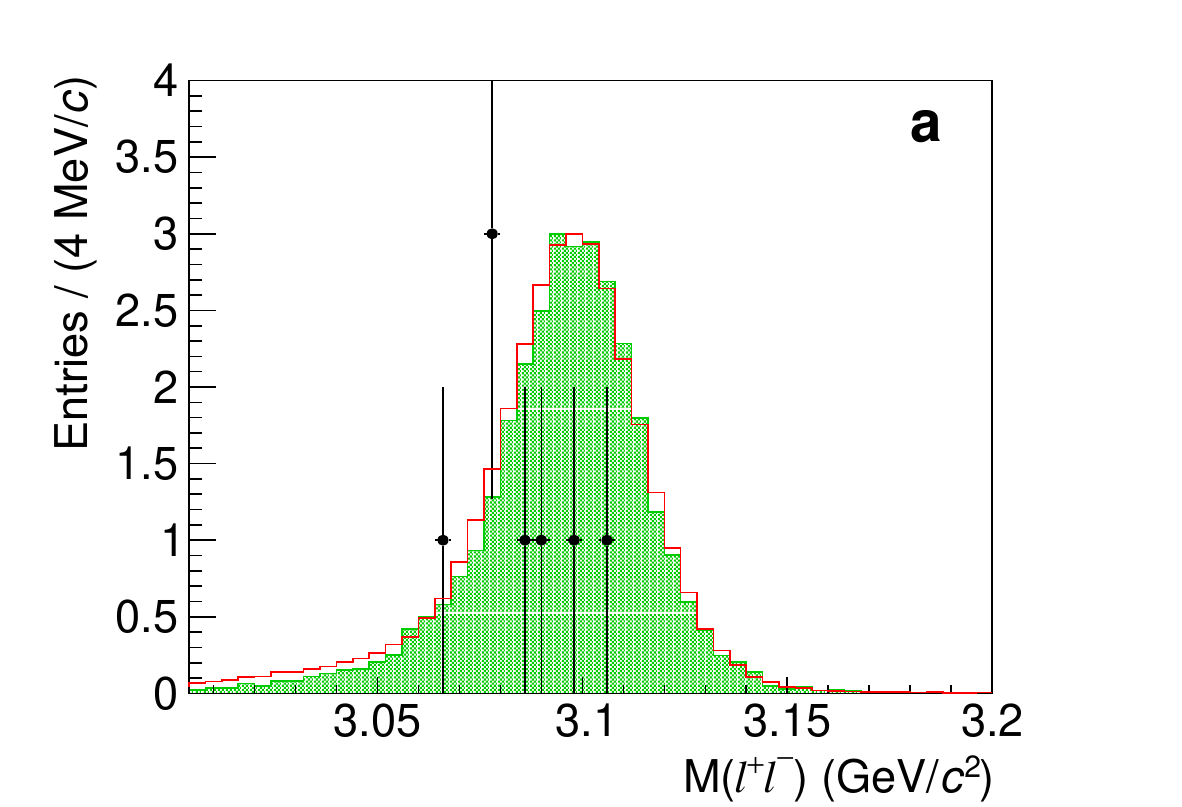}
  \includegraphics[width=0.42\textwidth]{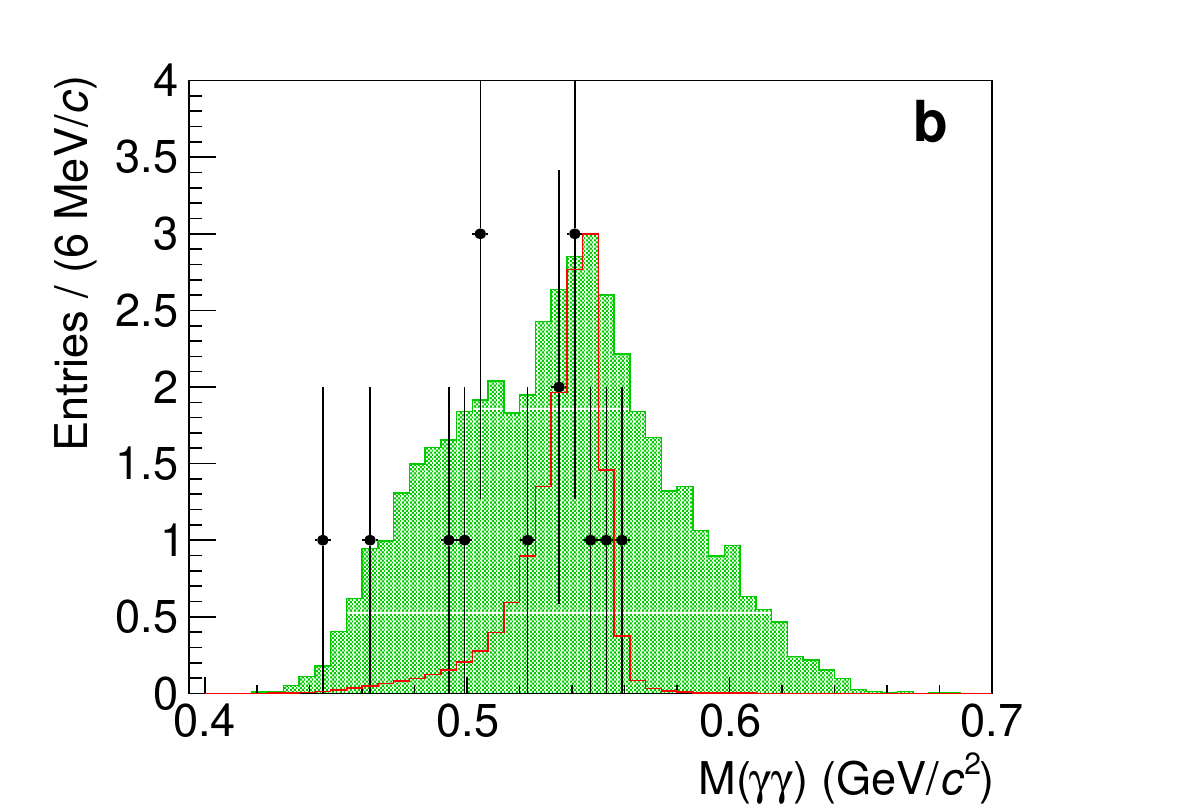}
  \caption{Invariant mass distributions  for (a) $\jpsi \to l^+l^-$ and (b) $\eta \to \gamma\gamma$ candidates at $\sqrt{s} = \SI{4.416}{GeV}$. Only events are chosen where both $\eta$s decay into $\gamma\gamma$.
           Black dots with error bars represent  data (the number of $\eta$ candidates is twice the number of $\jpsi$ candidates), red histograms  the signal MC sample, and the green area  the background MC sample. The scaling of the MC sample is arbitrary.}
  \label{fig:masses}
\end{figure}

The number of reconstructed events, $N_{\rm data}$, for each data sample is listed in Table~\ref{tab:xsec}. Only a few events per energy point survive the overall selection.
Figure~\ref{fig:masses} shows the invariant mass distributions
of $\ell^+\ell^-$, $M(\ell^+\ell^-)$, and $\gamma\gamma$, $M(\gamma\gamma)$, at $\sqrt{s} = \SI{4.416}{GeV}$, 
 after the selection and background suppression described above for the case where both $\eta$s decay into $\gamma\gamma$. The data distribution roughly follows the distribution of the background MC sample.

\begin{figure}[htbp]
  \centering
  \includegraphics[width=0.4\textwidth]{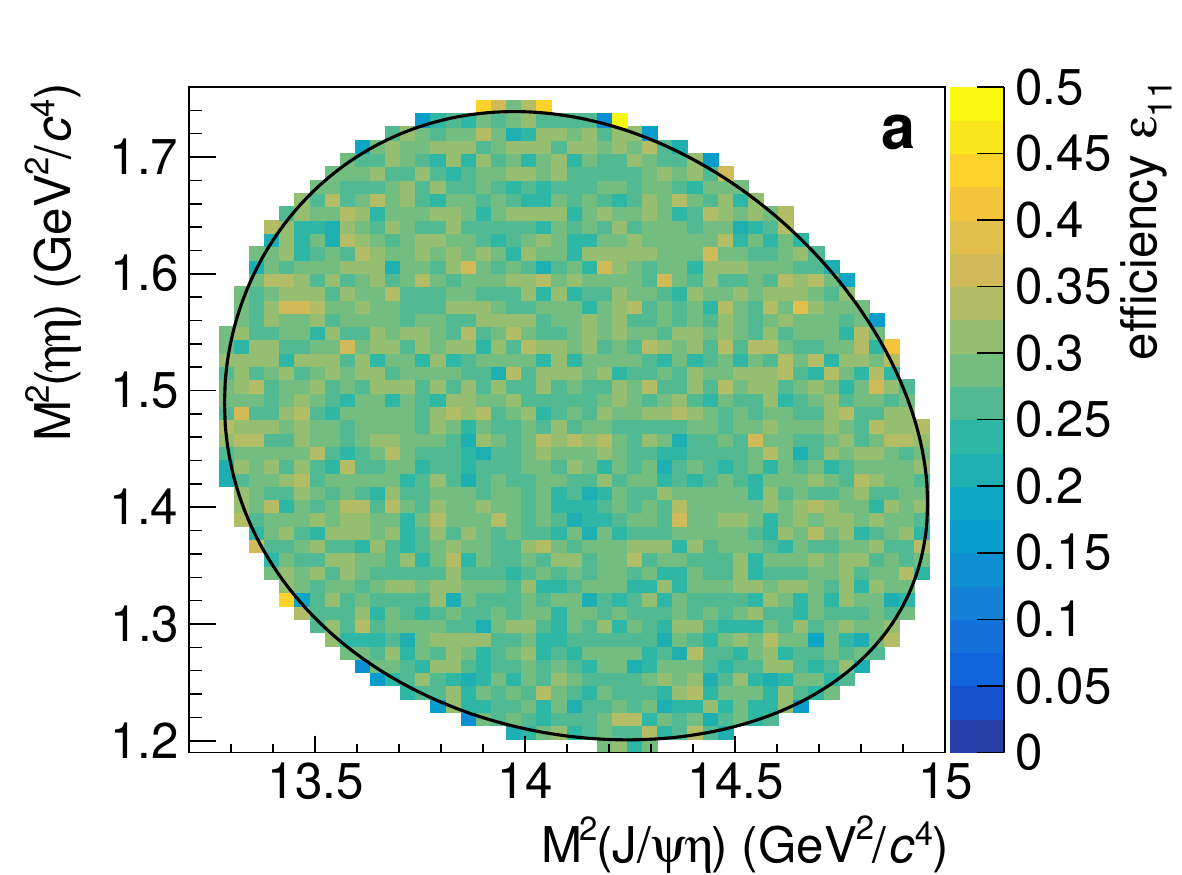}
  \includegraphics[width=0.4\textwidth]{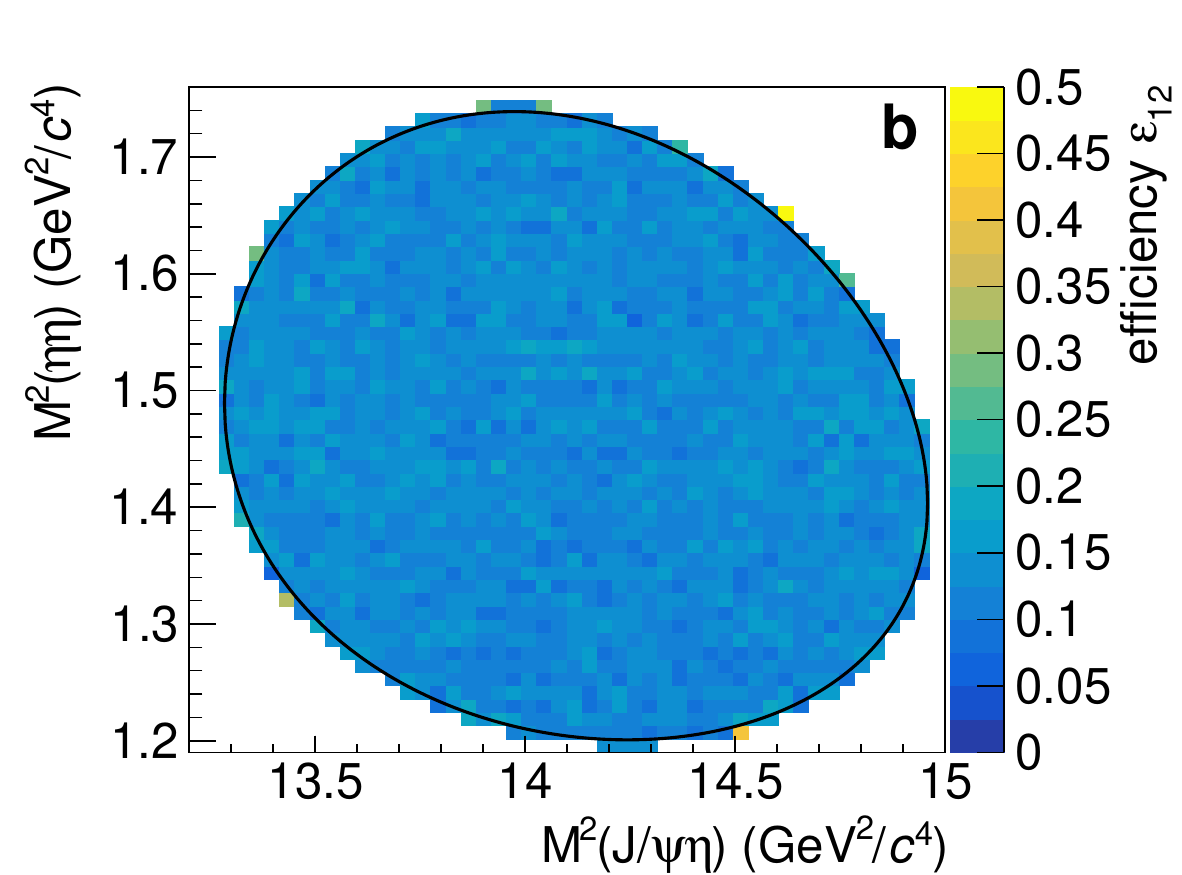}
  \includegraphics[width=0.4\textwidth]{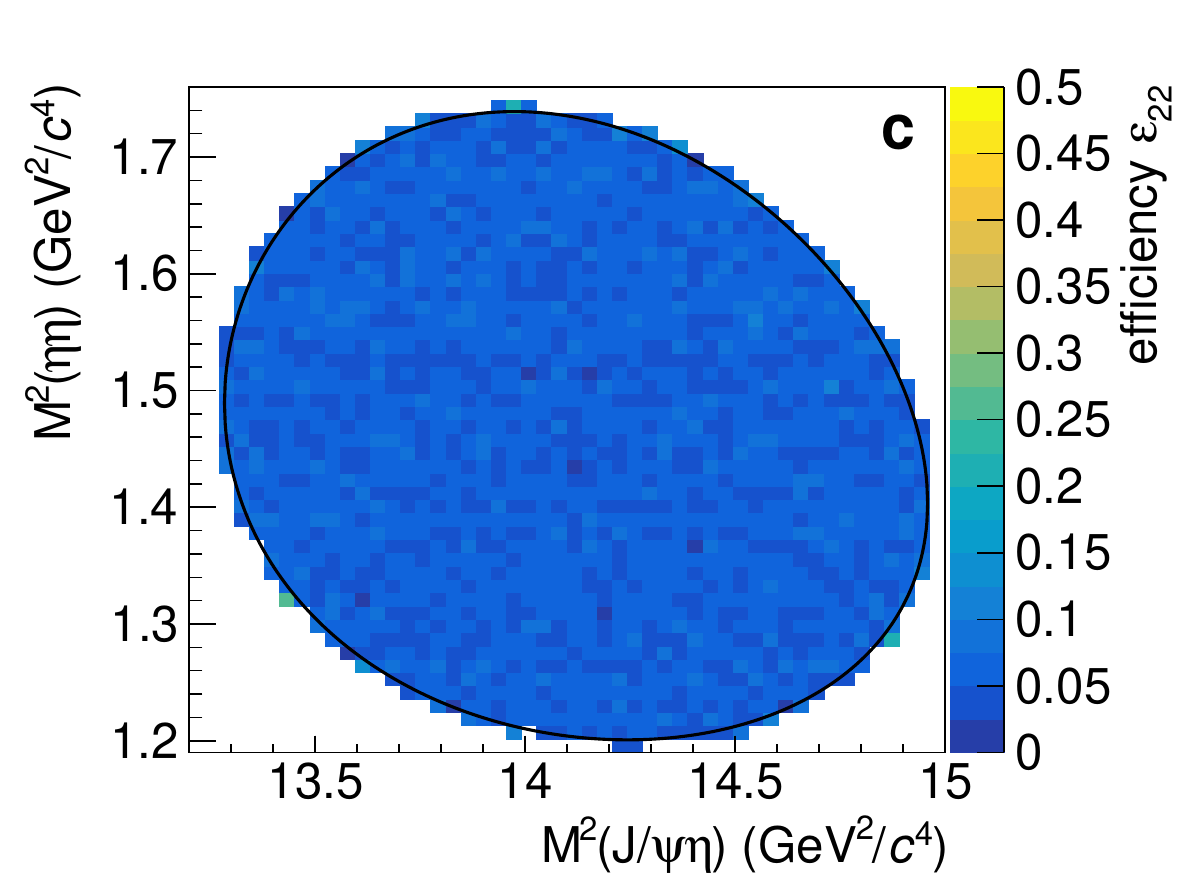}
  \caption{Two-dimensional efficiency distributions as a function of the squared invariant mass $M^2(\eta\eta)$ and $M^2(\jpsi \eta)$ at $\sqrt{s} = \SI{4.416}{\GeV}$ for (a) $\epsilon_{\rm 11}$ with $\eta\eta \to 4\gamma$, (b) $\epsilon_{\rm 12}$ with $\eta\eta \to 2\gamma\,\pip\pim\piz$ and (c) $\epsilon_{\rm 22}$ with $\eta\eta \to 2\left(\pip\pim\piz\right)$. }
  \label{fig:eff4420}
\end{figure}

The reconstruction efficiency $\epsilon$ gives the probability that a signal event is detected and survives the whole selection process. Since in this analysis two different $\eta$ decays are considered, three different reconstruction efficiencies, $\epsilon_{\rm 11}$, $\epsilon_{\rm 12}$ and $\epsilon_{\rm 22}$, are determined for each CM energy separately, where index 1 corresponds to $\twogg$ and index 2 to $\threepi$. The efficiency values are listed in Table~\ref{tab:xsec}. Figure~\ref{fig:eff4420} shows the two-dimensional (2D) distributions of the efficiencies as a function of the squared invariant mass of the $\eta\eta$ and of the $\jpsi\eta$ subsystems at $\sqrt{s} = \SI{4.416}{\GeV}$. All efficiency distributions are flat with no insensitive areas. The efficiencies for the $\twogg$ channel are higher  than for the $\threepi$. 

The observed cross section $\sigma^{\rm obs}$ for the reaction channel
$\elp \elm \to \eta\eta\jpsi$ is calculated as

\begin{equation}
 \sigma^{\rm obs} = \frac{N_{\rm data} - N_{\rm bkg}}{ \mathcal{L} \cdot \sum_{i}\limits \mathcal{B}_{i}\, \epsilon_{i} },
\end{equation}

\noindent
where $N_{\rm data}$ is the number of reconstructed events while $N_{\rm bkg}$ is the number of estimated background events scaled to the time-integrated luminosity $\mathcal{L}$ of the corresponding data sample; $\mathcal{B}_{i}$ and $\epsilon_{i}$ are the branching fractions and reconstruction efficiencies for the $i$ subsequent decays of the $\jpsi$ and the $\eta$. The branching fractions used for this analysis are $\mathcal{B}(\jpsi \to \elp\elm) = (5.971 \pm 0.032)$\%, $\mathcal{B}(\jpsi \to \mup\mum) = (5.961 \pm 0.0323)$\%, $\mathcal{B}(\twogg) = (39.36 \pm 0.18 )$\% and $\mathcal{B}(\threepi) = (23.02 \pm 0.25)$\%~\cite{ParticleDataGroup:2024cfk}.

For the determination of the statistical uncertainty of the observed cross section, the Poissonian confidence interval on $N_{\rm data}$ and $N_{\rm bkg}$ is used, and the uncertainty is calculated via Gaussian error propagation. 
The systematic uncertainty of the observed cross section consists of two parts. The first part $\Delta\sigma_{\rm \mathcal{L}red}$ contains the uncertainties related to the efficiency, e.g. the track reconstruction efficiency and the efficiency of the kinematic fit, and it is described in detail in section \ref{sec:systematics}. The second part is the systematic uncertainty from the background estimation $\Delta\sigma_{\rm bkg}$, which is obtained by varying the cross sections of the background channels within their statistical uncertainties and recalculating the number of background events.
Both systematic uncertainties are taken into account via Gaussian error propagation to determine their effect on the observed cross section. The results of the observed cross sections together with both systematic uncertainties are given in Table~\ref{tab:xsec}. 

\begin{turnpage}
\begin{table*}[htbp]
  \centering
	\addtolength{\tabcolsep}{1.5pt}
  \caption{Observed cross sections $\sigma^{\rm obs}$, Born cross sections $\sigma^{\rm Born}$ and upper limits at 90\% CL $\sigma^{\rm UL}$ of the exclusive method for the reaction channel $\elp \elm \to \eta\eta\jpsi$ as a function of the CM energy $\sqrt{s}$. The first uncertainty of the observed/Born cross sections $\sigma^{\rm obs}$ and $\sigma^{\rm Born}$ is statistical, and the second one systematic. The values needed for the $\sigma^{\rm obs}$ calculation: the time-integrated luminosity $\mathcal{L}$, the efficiencies $\epsilon_{\rm 11}$, $\epsilon_{\rm 12}$ and $\epsilon_{\rm 22}$ for the three cases of $\eta$-combinations, the number of reconstructed events $N_{\rm data}$, and the remaining background events $N_{\rm bkg}$ (normalized to the measured time-integrated luminosity). In addition, the single contributions to the systematic uncertainties for the observed cross section $\Delta\sigma_{\rm \mathcal{L}red}$ and $\Delta\sigma_{\rm bkg}$ originating from the reduced luminosity $\mathcal{L}_{\rm red}$ and the background contributions are listed separately. $\mathcal{SS}$ is the statistical significance of $\sigma^{\rm Born}$. The values needed for the $\sigma^{\rm UL}$ calculation: the systematic uncertainty $\Delta\mathcal{L}_{\rm red}$ and the ISR correction factor combined with the vacuum polarization factor $\frac{(1+\delta)}{|1-\Pi|^{2}}$. }
  \label{tab:xsec}
{\footnotesize
  \begin{tabular}{crc|ccccc|ccr|rc|ccc}
    \hline\hline  \\[-1em]
    $\sqrt{s}~(\si{GeV})$ & \multicolumn{2}{r|}{$\mathcal{L}~(\si{\pb^{-1}})$} & $\epsilon_{\rm 11}$ & $\epsilon_{\rm 12}$ & $\epsilon_{\rm 22}$ & $N_{\rm data}$ & $N_{\rm bkg}$ & $\Delta\sigma_{\rm \mathcal{L}red}~(\si{\pb})$ & $\Delta\sigma_{\rm bkg}~(\si{\pb})$ & $\sigma^{\rm obs}~(\si{\pb})$\;\;\;\;\;\; & $\sigma^{\rm Born}~(\si{\pb})$\;\;\;\;\;\; & $\mathcal{SS}$~($\sigma$)
    & $\Delta\mathcal{L}_{\rm red}~(\si{\per\pb})$ & $\frac{(1+\delta)}{\left|1-\Pi\right|^{2}}$ & $\sigma^{\rm UL}~(\si{\pb})$ \\
    \hline \\[-1em]
\input{crosssection}
    \\\hline\hline
  \end{tabular}
  }
\end{table*}
\end{turnpage}

The Born cross section
\begin{equation}
 \sigma^{\rm Born} = \sigma^{\rm obs} \cdot {\frac{|1-\Pi|^{2}}{(1+\delta)}}
\end{equation}
is determined by dividing the observed cross section with the combined ISR and vacuum polariszation correction factor $\frac{(1+\delta)}{|1-\Pi|^{2}}$. The determination of the correction factor is described in detail in section~\ref{sec:ISR}. The results of the Born cross sections are shown in Fig.~\ref{fig:xsec-over-e} and listed together with the correction factors and also the significance in Table~\ref{tab:xsec}. The quoted statistical significance is based on the binomial assumption  $Z_{\rm Bi}$, from Ref.~\cite{COUSINS2008480}, and does not include systematic uncertainties.

Almost all Born cross sections up to 4.6\,GeV CM energy are statistically not significant and are compatible with zero within their statistical and systematic uncertainties. Only in three cases an evidence could be claimed with more than 3$\sigma$ significance.  Above 4.6\,GeV, the significance is higher in general. At two CM energies a high significance of 5.7$\sigma$ (4.682\,GeV) and 8.9$\sigma$ (4.750\,GeV) and at three CM energies an significance of more than 3$\sigma$ is observed. However, some of the background channels are not measured yet above 4.6\,GeV and thus the estimation of the background contribution is more uncertain. Therefore, no observation or evidence is claimed and the upper limits at 90\% confidence level (CL) of the Born cross sections are determined.

\subsection{Intermediate states}
\begin{figure} [htbp]
  \centering
  \includegraphics[width=0.42\textwidth]{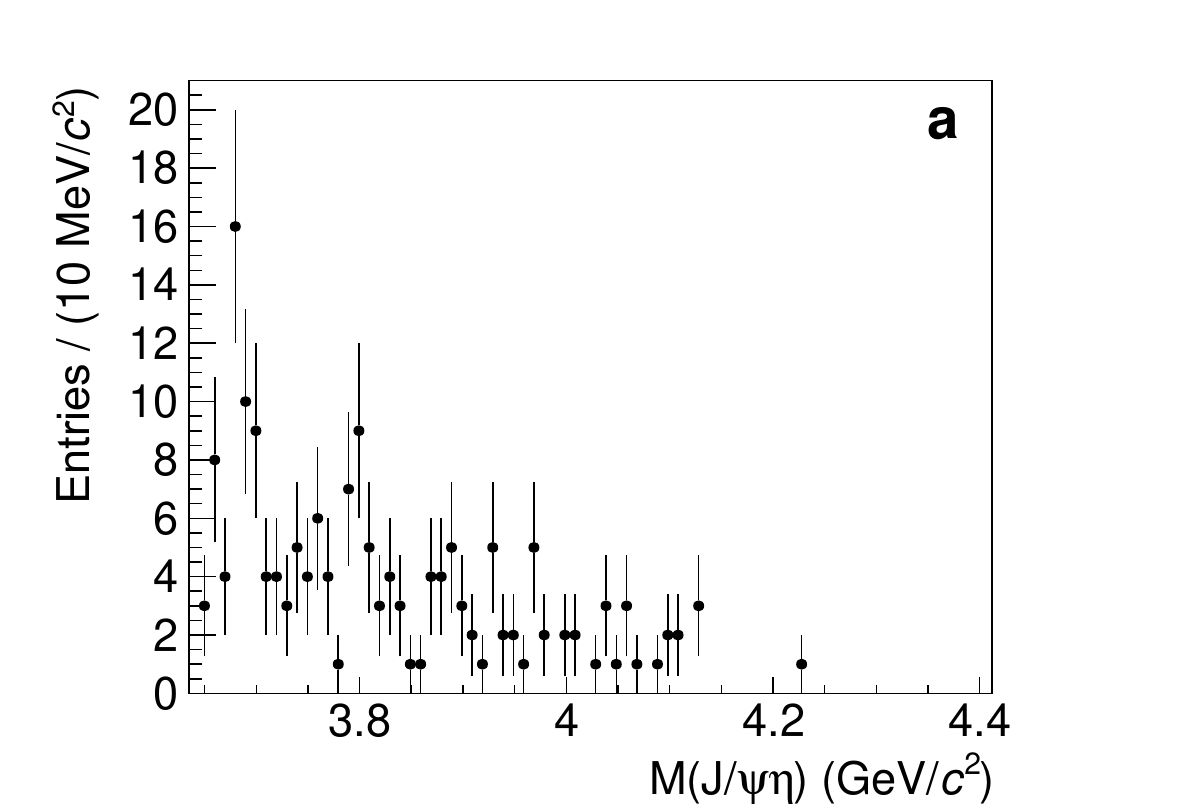}
  \includegraphics[width=0.42\textwidth]{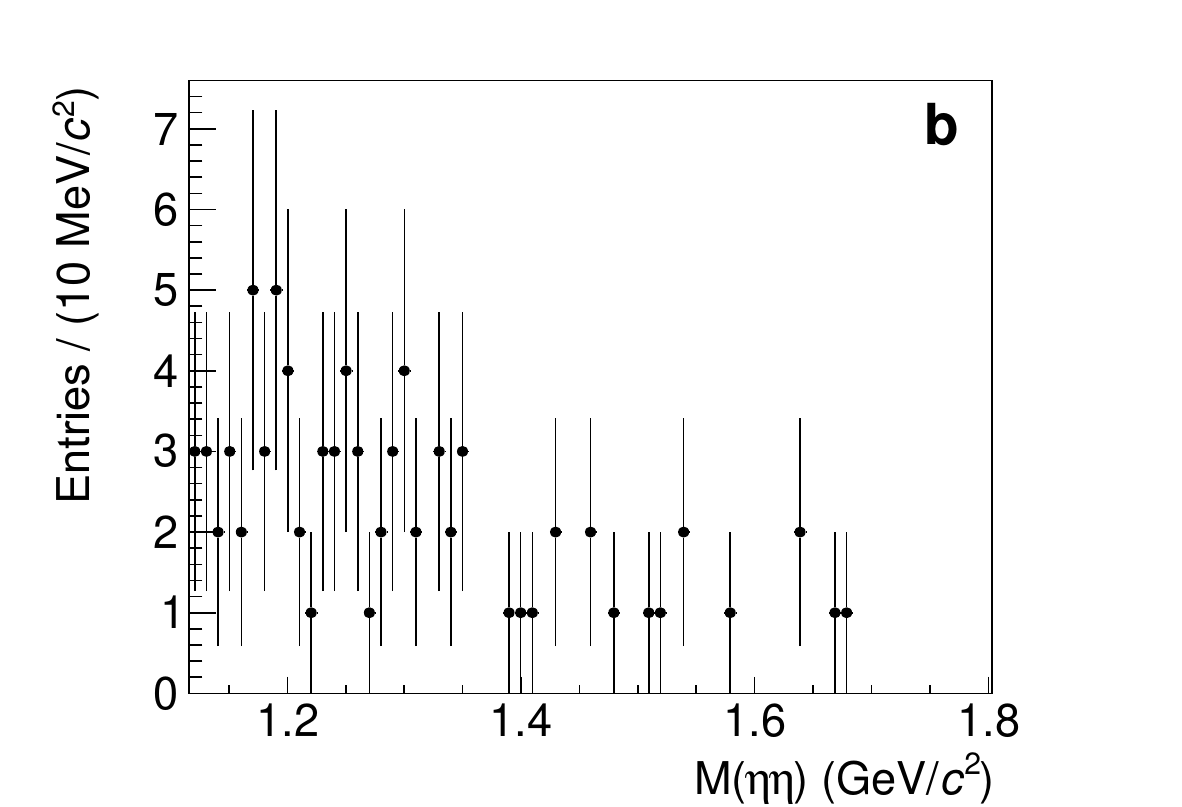}
  \caption{Distributions of (a) $M(\jpsi\eta)$ (two entries per event) and (b) $M(\eta\eta)$ (one entry) summed over all energies.}
  \label{fig:masses_all}
\end{figure}

We investigate the presence of possible intermediate resonances by studying the $\jpsi\eta$ and $\eta\eta$ invariant mass distributions. Figure~\ref{fig:masses_all} shows the $M(\jpsi\eta)$ and $M(\eta\eta)$ distributions,
summing up all energies and assuming that all reconstructed events are signal events.

In the $M(\jpsi\eta)$ distribution (Fig.~\ref{fig:masses_all}a), a clear signal of the reaction channel $\elp\elm\to\eta\psip$ with $\psip\to\jpsi\eta$ is visible, and two additional enhancements appear around 3.8 and 3.9 GeV/$c^{2}$.  No enhancement can be seen in the 
$M(\eta\eta)$ distribution (Fig.~\ref{fig:masses_all}b). Nevertheless, due to the limited statistics, no conclusion can be drawn on the existence of intermediate states.

\subsection{Systematic uncertainty}
\label{sec:systematics}
As the number of observed events in data is quite small, the upper limit will be determined using the method described in Ref.~\cite{ref:Rolke2004mj}. For this method, the systematic uncertainties are inserted by increasing the uncertainty (width of a Gaussian function) of the reconstruction efficiency. 

Uncertainties from the time-integrated luminosity and the branching fractions  contribute to the total systematic uncertainty, but do not affect the reconstruction efficiencies themselves.
Therefore, including these numbers in the width of the Gaussian distribution would overestimate the overall systematic uncertainty.
To avoid this problem, the ``reduced luminosity" $\mathcal{L}_{\rm red}$ is defined and this value is used as the mean of the Gaussian distribution in the calculation of the upper limit:

\begin{equation}
  \mathcal{L}_{\rm red} = \mathcal{L} \cdot \sum_{i}\limits \mathcal{B}_{i}\, \epsilon_{i}\,\alpha_{i}\,.
  \label{equ:lred}
\end{equation}
where $\mathcal{L}$ is the time-integrated luminosity of the corresponding data sample; $\mathcal{B}_{i}$ and $\epsilon_{i}$ are the branching fractions and reconstruction efficiencies for the $i$ subsequent decays of the $\jpsi$ and the $\eta$ and the correction factor $\alpha_{i}$. For the sources of systematic uncertainties which are associated with a certain value, e.g. the branching fractions or the reconstruction efficiency, their nominal value is taken. The correction factor $\alpha_{i}$ is set to 1. 

To determine the systematic uncertainty of each source from the formula above, $\mathcal{L}_{\rm red}$ is recalculated 1000 times after the corresponding value is varied by drawing a number within a Gaussian function with the width corresponding to the uncertainty of the investigated value. The root mean square of the $\mathcal{L}_{\rm red}$ distribution is taken as the systematic uncertainty. 

For the determination of the uncertainty caused by the systematic uncertainties of the track and photon reconstruction (which is given by 1\% per track and 1\% per photon), one number for tracks $\Delta\alpha_{tr}$ and separately one number for photons $\Delta\alpha_{ph}$ is drawn from a Gaussian distribution with the width of 1\%. Then the number $n_i$ of charged tracks and the number $m_i$ of photons is used to change the correction factor 
 $\alpha_i = \left( 1 + \Delta\alpha_{tr}\right)^{n_i} \cdot \left( 1 + \Delta\alpha_{ph}\right)^{m_i}$, before recalculating $\mathcal{L}_{\rm red}$. 

The following uncertainties are estimated in a different way: for the $\piz\piz$ veto, the cut is varied by $\pm10$ units of the $\chi^{2}$-difference; the largest difference to the nominal value is taken as the uncertainty.
For the kinematic fit, the cut on $\chi^2$ is varied by 5\%; the largest difference to the nominal value is taken as the uncertainty.
The uncertainty of the reconstruction efficiency $\epsilon$ is deduced from the statistical uncertainty due to the size of the signal MC samples.

The results for each CM energy are summarized in Table~\ref{tab:excl:syserr}. 

\begin{table}[htbp]
  \center
  \caption{The absolute systematic uncertainty $\Delta\mathcal{L}_{\rm red}$ for the upper limit determination of the reaction channel $\elp\elm\to\eta\eta\jpsi$, as described in the text. The range reflects the variation between the different CM energies. For comparion with the semi-inclusive method also the relative systematic uncertainty $\Delta\mathcal{L}_{\rm red}/\mathcal{L}_{\rm red}$ is indicated.}
  \begin{tabular}{cc}
    \hline \hline  \\[-1em]
    Source & Systematic uncertainty \\
    \hline  \\[-1em]
    Luminosity							&\SIrange{0.008}{0.085}{\per\pb}\\
    Branching fractions					&\SIrange{0.007}{0.097}{\per\pb}\\
    Track \& photon reconstruction    	&\SIrange{0.025}{0.381}{\per\pb}\\
    $\piz\piz$ veto  				&\SIrange{0.001}{0.034}{\per\pb}\\
    Kinematic fit      					&\SIrange{0.004}{0.046}{\per\pb}\\
    Efficiency			  				&\SIrange{0.003}{0.034}{\per\pb}\\
    \hline  \\[-1em]
    $\Delta\mathcal{L}_{\rm red}$                					&\SIrange{0.031}{0.413}{\per\pb}\\
    $\Delta\mathcal{L}_{\rm red}/\mathcal{L}_{\rm red}$     					&\SIrange{3.62}{4.03}{\%}\\
    \hline \hline
  \end{tabular}
  \label{tab:excl:syserr}
\end{table}


\subsection{ISR correction and Vacuum polarization}
\label{sec:ISR}
For the calculation of the upper limits corresponding to the Born cross sections, the correction of ISR and vacuum polarization needs to be determined. 

\begin{figure}[htbp]
    \center
    \includegraphics[width=0.45\textwidth]{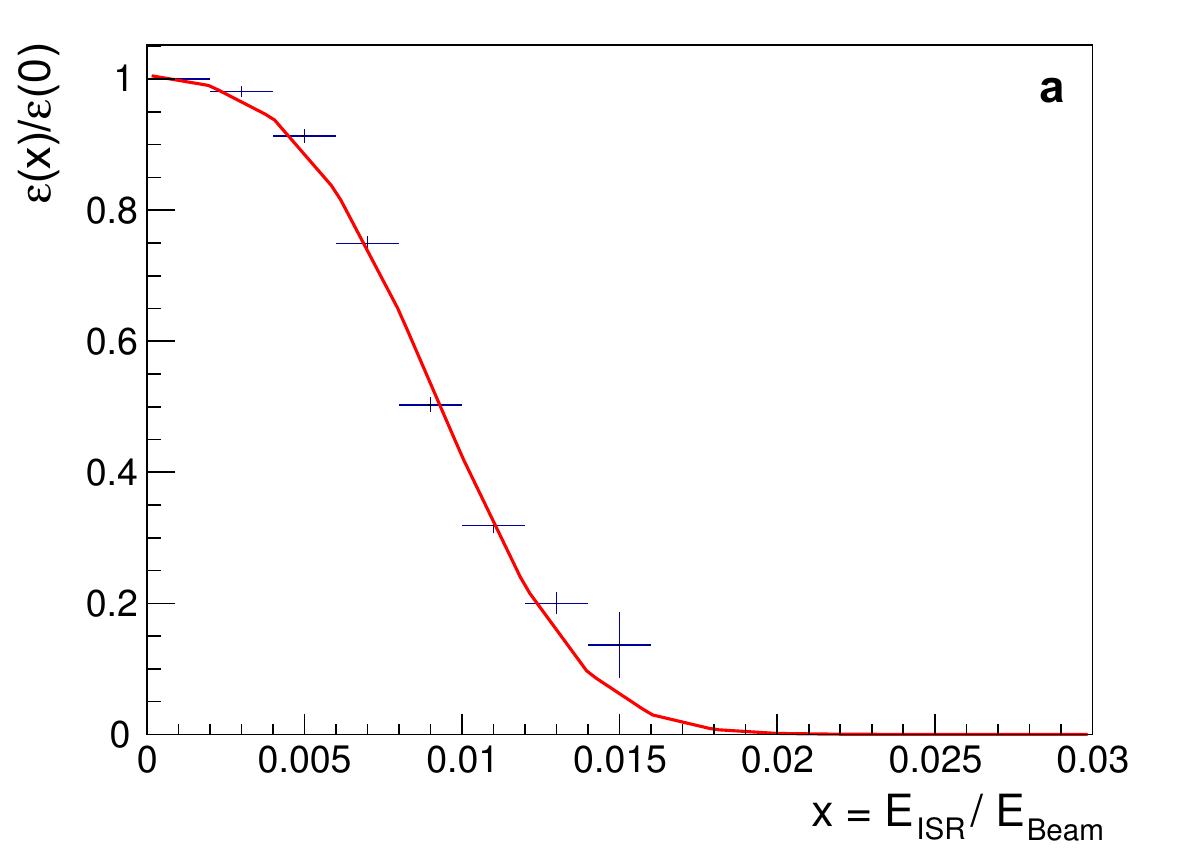}
    \includegraphics[width=0.45\textwidth]{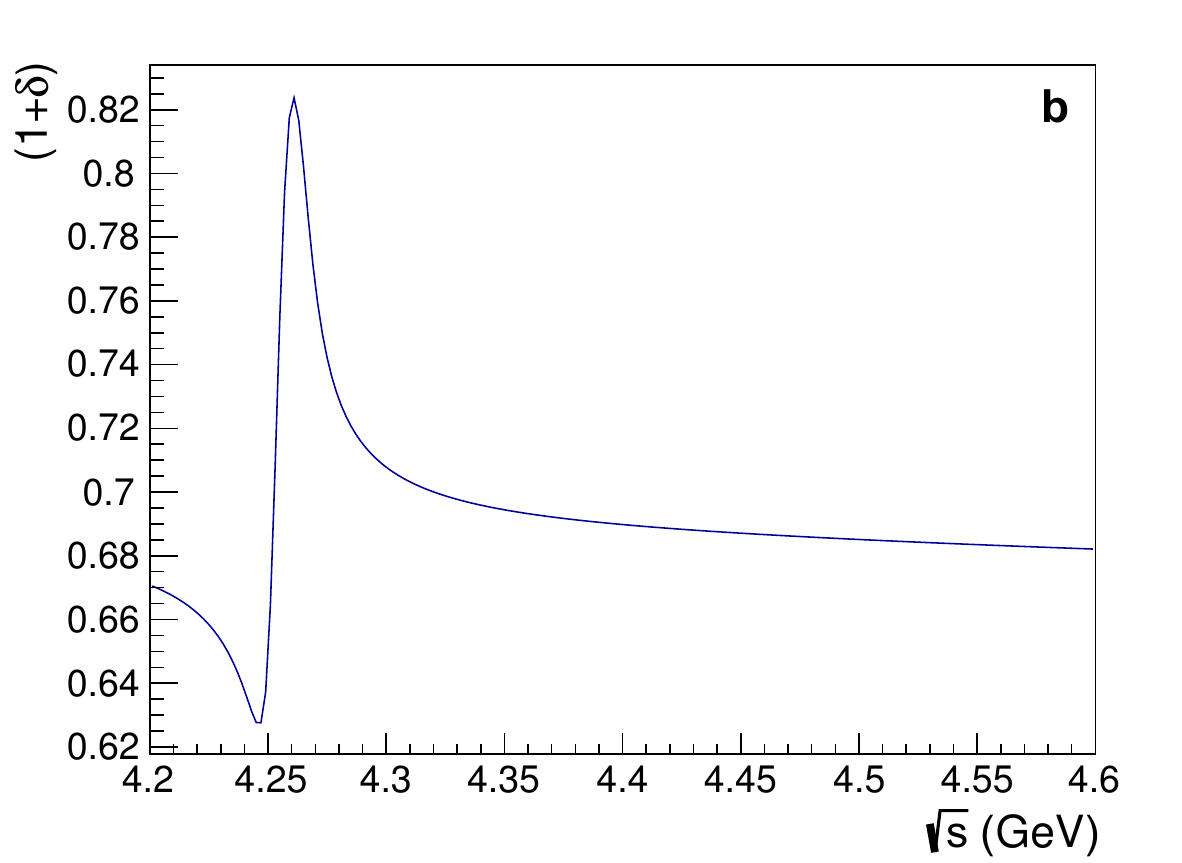}
\caption{ISR correction for the reaction channel $\elp \elm \to \eta \eta \jpsi$, $\eta\eta \to 4\gamma$, at $\sqrt{s}= \SI{4.360}{\MeV}$. (a) Normalized reconstruction efficiency versus $E_{\rm ISR}/E_{\rm beam}$, where the red curve represents the fit with the error function;
(b) dependence of the ISR correction factor $(1+\delta)$ on $\sqrt{s}$, assuming a single narrow resonance with mass of 4.260 GeV/$c^{2}$ and width of $\SI{10}{\MeV}$.}
    \label{fig:isr_corr_bw}
\end{figure}

The number of observed events $N$ can be written as
\begin{equation}
N = \mathcal{L} \int \sigma (x) \epsilon (x) W(x) {\rm d}x,
\end{equation}
where $\mathcal{L}$ is the time-integrated luminosity, $x \equiv E_{\rm ISR}/E_{\rm beam}$, the quotient of the energy of the ISR photon divided by the energy of the beam, and $\sigma(x)$, $\epsilon(x)$ and $W(x)$ being the cross section, the efficiency and the radiator function determined for $x$~\cite{MONTAGNA199931,BESIIISYS:2014}.
After factorizing out the Born cross section $\sigma _{\rm 0}$ and the efficiency $\epsilon _{\rm 0}$ at $x=0$ (with vanishing energy of the ISR photon), this expression becomes
\begin{equation}
N = \mathcal{L} \sigma _{\rm 0} \epsilon _{\rm 0} \int \frac{\sigma (x)}{\sigma _{\rm 0}} \frac{\epsilon (x)}{\epsilon _{\rm 0}} W(x) {\rm d}x.
\end{equation}
The ISR correction factor $(1+\delta)$ is defined as
\begin{equation}
(1+\delta) \equiv \int \frac{\sigma (x)}{\sigma _{\rm 0}} \frac{\epsilon (x)}{\epsilon _{\rm 0}} W(x) {\rm d}x 
\end{equation}
so that
\begin{equation}
N = \mathcal{L} \sigma _{\rm 0} \epsilon _{\rm 0} (1+\delta).
\end{equation}
The correction factor is calculated by using numerical integration. First, the normalized efficiency $\epsilon (x)/ \epsilon _{\rm 0}$ as a function of $x$ is determined by using a signal MC sample produced including ISR. Figure~\ref{fig:isr_corr_bw}a shows the result for the CM energy of $\SI{4.360}{\GeV}$ with $\eta\eta \to 4\gamma$, together with an error function fitted to the results. For all final states and CM energies the error function is suitable to describe the  $\epsilon (x)/ \epsilon _{\rm 0}$ dependence.

The correction factor $(1+\delta)$ is strongly correlated to the energy dependence 
of the signal cross section, which is currently unknown. To obtain a
conservative upper limit, the lowest possible
$(1+\delta)$ is estimated, assuming a narrow resonance structure in the cross section with a mass of 4.260 GeV/$c^{2}$ and a width of $\SI{10}{\MeV}$. The resulting energy dependence of $(1+\delta)$ on the CM energy is shown in Fig.~\ref{fig:isr_corr_bw}b. 
Changing the position of the resonance results in a corresponding shift of the $(1+\delta)$ energy dependence, while the shape is nearly unchanged. The minimal value of   
the correction factor, $(1+\delta) = 0.67$, is conservatively used to set the
upper limits of the cross sections at all CM energies. 
The vacuum polarization is determined by using the tool based on alphaQED~\cite{FJegerlehner2017}.
The combined correction value is listed in Table~\ref{tab:xsec}.


\subsection{Upper limit determination}
\label{sec:upperlimit}

The upper limits on the cross sections are calculated following a frequentist procedure~\cite{COUSINS2008480,ref:Rolke2004mj},
using the definition
\begin{equation}
 \sigma^{\rm UL}~\equiv~\frac{N_{\rm UL}}{\mathcal{L}\;(1+\delta)\;\frac{1}{|1-\Pi(s)|^{2}}~\sum_{i}\limits \mathcal{B}_{i}\, \epsilon_{i}~}
 \label{equ::sigma_UL}
\end{equation}
with $N_{\rm UL}$ being the upper limit on the signal yield,
$\mathcal{L}$ the time-integrated luminosity, $(1+\delta)$ the ISR correction factor, $\frac{1}{|1-\Pi(s)|^{2}}$ the vacuum polarization correction factor, $\epsilon_i$ the reconstruction efficiency, and $\mathcal{B}_i$ the combined branching ratio $\mathcal{B}_i$ of the three different $\eta\eta$ final state combinations $i$ and including the branching ratios of $\mathcal{B}({\jpsi \to \elp \elm})$ and $\mathcal{B}({\jpsi \to \mup \mum})$.

The systematic uncertainties are taken into account by assuming a Gaussian-shaped uncertainty on the  
efficiency with a width equal to the total systematic uncertainty. The reduced luminosity $\mathcal{L}_{\rm red}$ was defined in equation (\ref{equ:lred}) for this purpose. The upper limits on the cross sections are then determined by inserting $\mathcal{L}_{\rm red}$ in equation (\ref{equ::sigma_UL}):

\begin{equation}
 \sigma^{\rm UL}~\equiv~\frac{N_{\rm UL}}{\mathcal{L}_{\rm red}\;(1+\delta)\;\frac{1}{|1-\Pi(s)|^{2}}~}.
\end{equation}

\begin{figure}[htbp]
  \centering
  \includegraphics[width=0.5\textwidth]{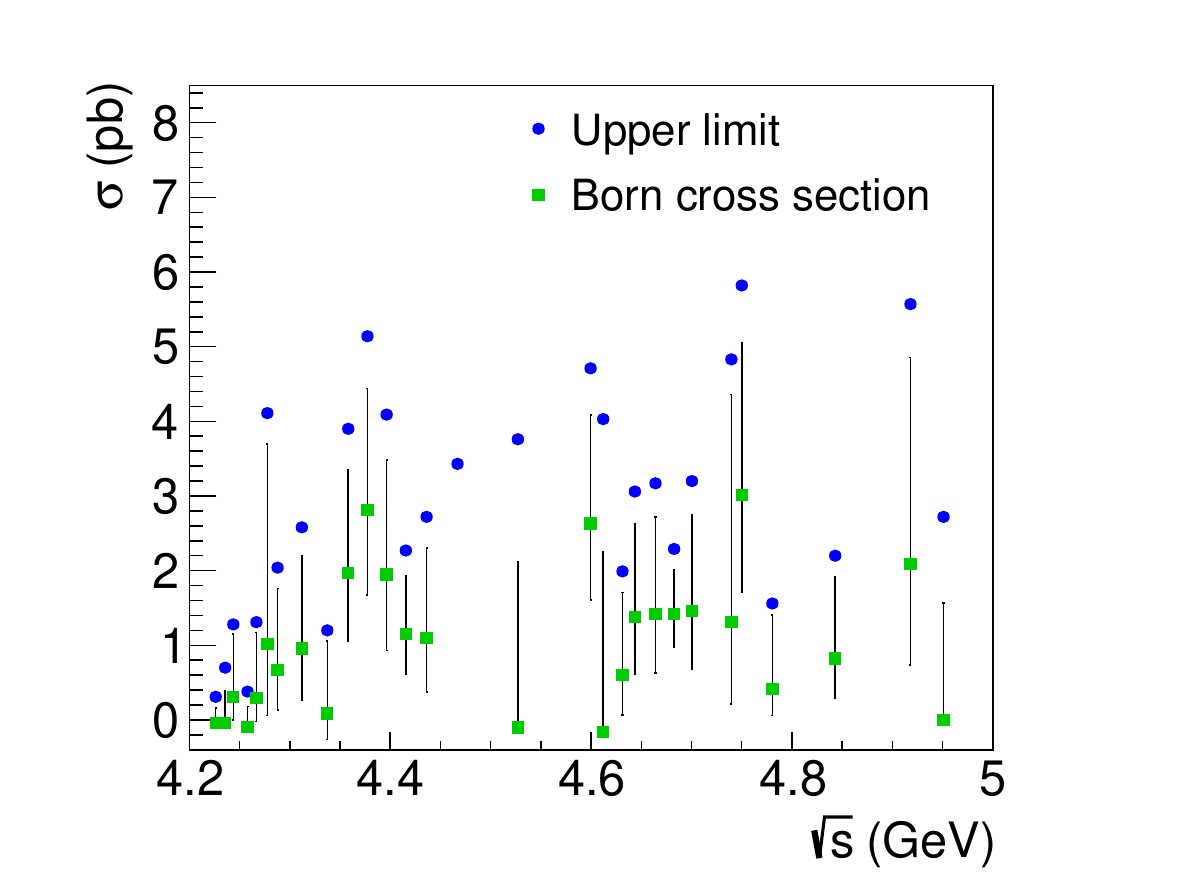}
  \caption{Born cross sections (green) and upper limits (blue) at the 90\% CL on the Born cross section as a function of the CM energy. For the Born cross section, only the statistical uncertainties are shown as error bars.}
  \label{fig:xsec-over-e}
\end{figure}

The results of the upper limits are given in Table 
\ref{tab:xsec}, together with the input values, and are shown in Fig.~\ref{fig:xsec-over-e}. The enhancement in the upper limits between $\SI{4.4}{}$ and $\SI{4.7}{GeV}$ depends on the low statistics of the corresponding data samples. 

\subsection{Summary --- Exclusive method}
In this section, the exclusive reconstruction of the complete event is used to search for the decay $\elp\elm\to\eta\eta\jpsi$ at the twenty-nine energy values with $\sqrt{s}$ between
4.226 and 4.950 GeV. At six data points evidence of the decay channel is found with more than 3$\sigma$ significance for the cross section determination. At two CM energies (4.682 with 5.7$\sigma$ and 4.750~GeV with 8.9$\sigma$) an observation could be claimed. Since only a few events are observed and due to the remaining background contributions, the statistical significance of most of the results is less than 3$\sigma$ (see Table~\ref{tab:xsec}). Therefore, the upper limit for each CM energy is determined at the 90\% confidence level. Both results are shown in Fig.~\ref{fig:xsec-over-e}, where no clear structure is visible.


\section{Semi-inclusive method}
\label{ref:in}
For this method, one $\eta$ has to decay to $\gamma\gamma$, while the second one is allowed to decay to anything. 

\subsection{Data analysis}

For the selection of the lepton candidates, the total momentum of the tracks is required to be greater than $0.95$~GeV/$c$.
Leptons with energy deposit in the calorimeter greater than 0.95~GeV are assigned as electrons, otherwise as muons. The number of photons ($N_{\gamma}$) is required to be greater than one.

To suppress potential background contributions and to improve the resolution, a two-constraint (2C) kinematic fit is performed,
imposing the nominal $\jpsi$ mass~\cite{ParticleDataGroup:2024cfk} on the invariant mass of the lepton pair, and constraining the missing mass to the $\eta$ mass~\cite{ParticleDataGroup:2024cfk}.

If more than one combination of lepton or photon pairs is found, the candidate with the lowest $\chisq$ of the kinematic fit is chosen.
The $\chisq$ of the kinematic fit is required to be less than 10 by maximizing the figure-of-merit $S/\sqrt{S+B}$ (FOM), where $S$ is the signal yield estimated from the signal MC samples and normalized by the previously measured cross sections, and $B$ is the background yield estimated by the background MC samples and normalized according to the luminosity. The FOM is estimated in the signal region, which is defined as a recoil mass window of $\jpsi\eta$ $RM(\jpsi\eta) \in (533.6,\, 561.6)$~MeV/$c^{2}$.

For the $\mu\mu$-mode, we require the number of the track hit layers in the MUC to be greater than 6 to suppress the misidentification of pions as muons.
To suppress the main background contribution when the photon candidates do not originate from $\eta$,
we require that the ratio of absolute $\Delta E_{\gamma}$ to the missing momentum
is less than a FOM optimized value 0.95, where $\Delta E_{\gamma}$ is the energy difference of the two reconstructed photons.

\begin{figure}[htbp]
	\centering
	\includegraphics[width = 0.43\textwidth]{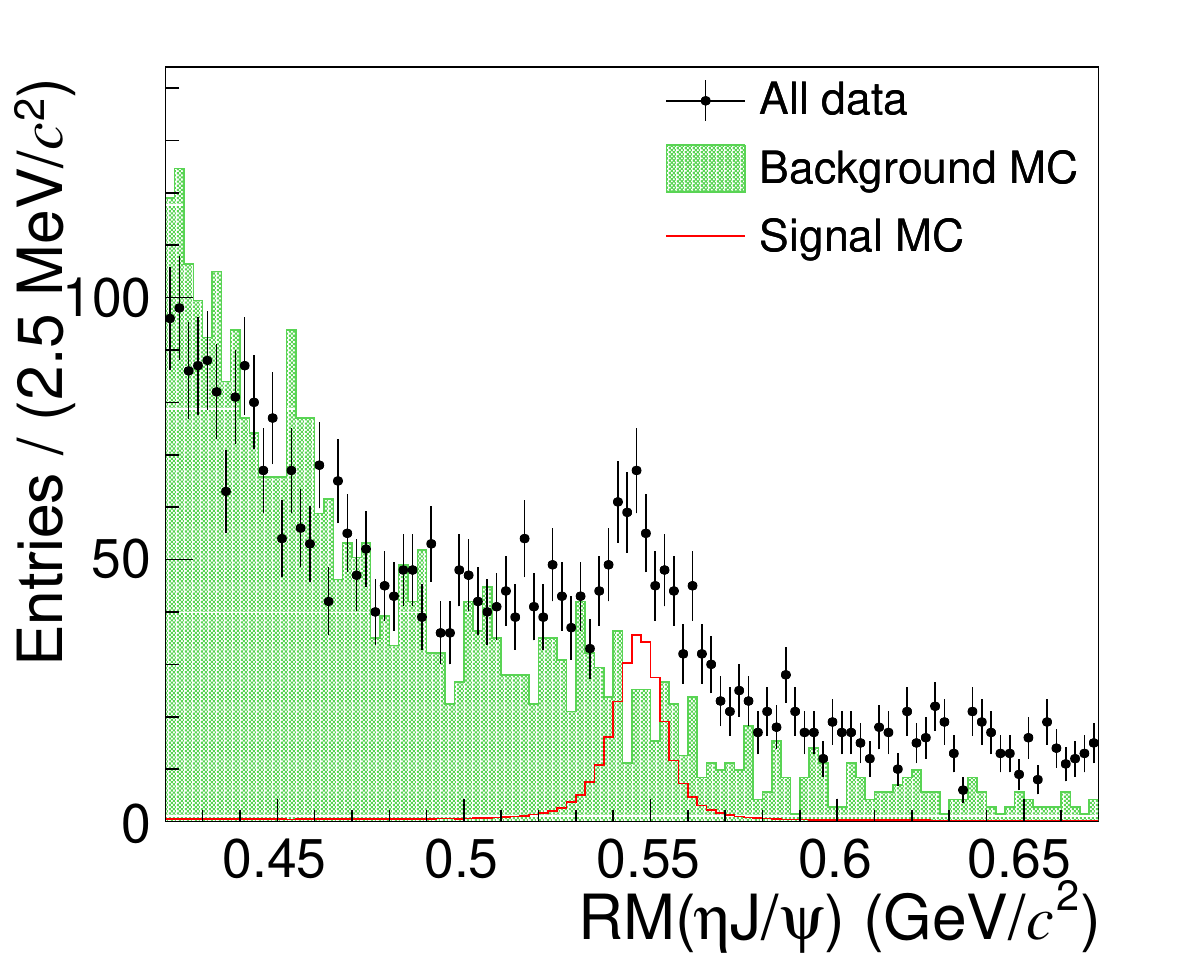}
	\caption{Distribution of the $\jpsi\eta$ recoil mass. The dots with error bars are the sum of all data samples, the green filled histogram is the sum of all background MC samples, and the red histogram is the sum of all signal MC samples, normalized using the measured cross sections.}
	\label{fig:recmetajpsi}
\end{figure}

Figure~\ref{fig:recmetajpsi} shows the $\jpsi\eta$ recoil mass distribution
after all the aforementioned event selection criteria, summing up the data samples and background MC samples at all CM energies.
A significant enhancement around the $\eta$ mass can be seen in the data samples. Looking at the distribution of the background MC samples no peaking background is found. It can also be seen, that the background MC is not describing the recoil mass distribution correctly at higher masses. This is no surprise, since only very few measurements exist in this energy range and the ratios used for the background MC production are mainly arbitrary. However, it indicates the main background channels which should be taken into account for the further analysis.  
The dominant background channels are $\elp \elm \to \gamma_{\rm ISR} \psip$ ($\psip \to\piz\piz\jpsi$)
and $\elp \elm \to \piz \piz \jpsi$, where $\piz$ decays into a photon pair, and $\jpsi$ decays into a lepton pair.

\subsection{Intermediate states}

\begin{figure}[htbp]
  \centering
  \begin{overpic}[width = 0.43\textwidth, trim={0.0 0.1cm 0.0 0.0}]{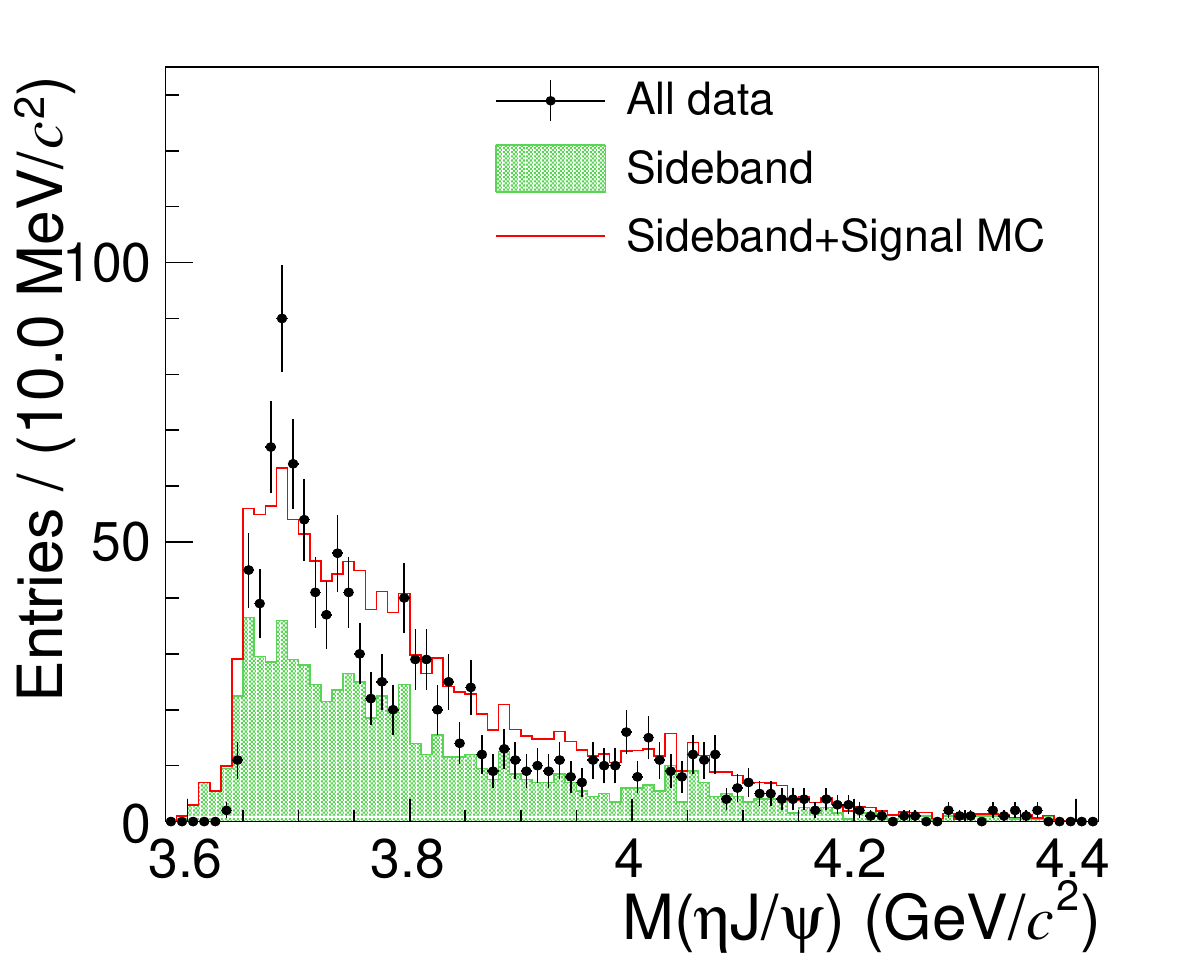}
		\put(87, 69){\bf\large a}
	\end{overpic} \\
	\begin{overpic}[width = 0.43\textwidth, trim={0.0 0.6cm 0.0 0.0cm}]{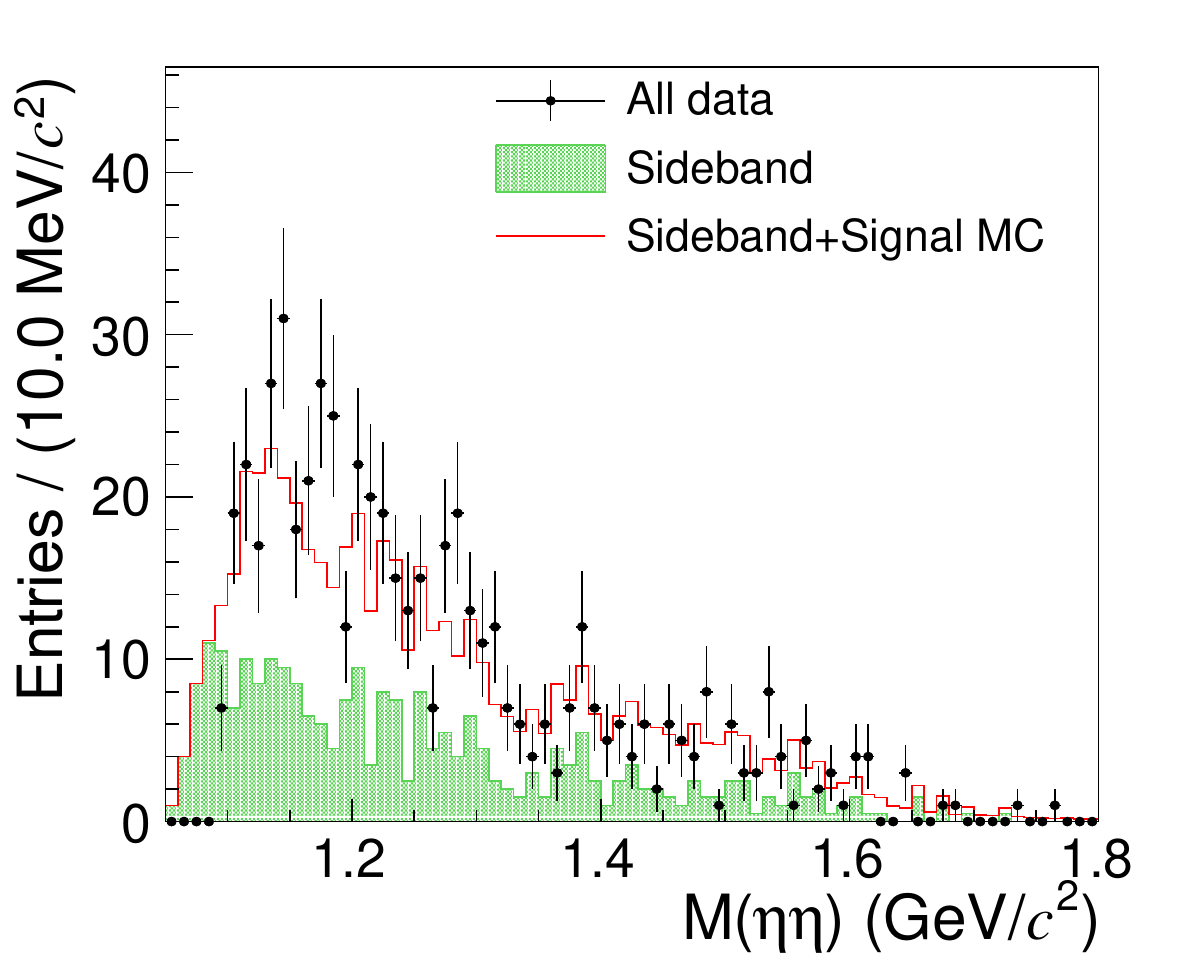}
		\put(87, 67){\bf\large b}
	\end{overpic}
  \caption{Invariant mass distributions of (a) $M(\jpsi\eta)$ (two entries per event) and (b) $M(\eta\eta)$. The dots with error bars are the sum of all data samples, the green filled histograms are background contributions from sidebands, and the red histograms are the sum of sidebands and signal MC samples, where the signal MC samples are normalized using the measured cross sections.}
  \label{fig:metajpsi}
\end{figure}

To investigate the presence of potential intermediate states in the reaction channel $\elp \elm \to \eta\eta\jpsi$, the invariant mass distributions of $\jpsi\eta$ and $\eta\eta$ 
are shown in Fig.~\ref{fig:metajpsi}. The signal MC distributions are normalized to the measured Born cross section at each energy point.

The $\eta$-mass sidebands of the recoil mass $RM(\eta\jpsi)$, between $(495.6\,-\,523.6)$ MeV/$c^{2}$ and $(571.6\,-\,599.6)$ MeV/$c^{2}$,
are used to describe the background distribution in data. Comparing all the data samples with the sum of the normalized sidebands and the signal MC distribution, 
(see Fig.~\ref{fig:metajpsi}a) we find clearly the reaction channel $\elp\elm\to\eta\psip$ with $\psip\to\jpsi\eta$.
In Fig.~\ref{fig:metajpsi}b there may be an indication of the $f_{2}(1270)$.
Except for the $\psip$ and the $f_{2}(1270)$ resonances, we do not find any other clear indication for peaking structures
in both distributions with the current statistics.

\subsection{Born cross section}

\begin{figure}[htbp]
	\centering
	\begin{overpic}[width = 0.43\textwidth, trim={0.0 0.1cm 0.0 0.0}]{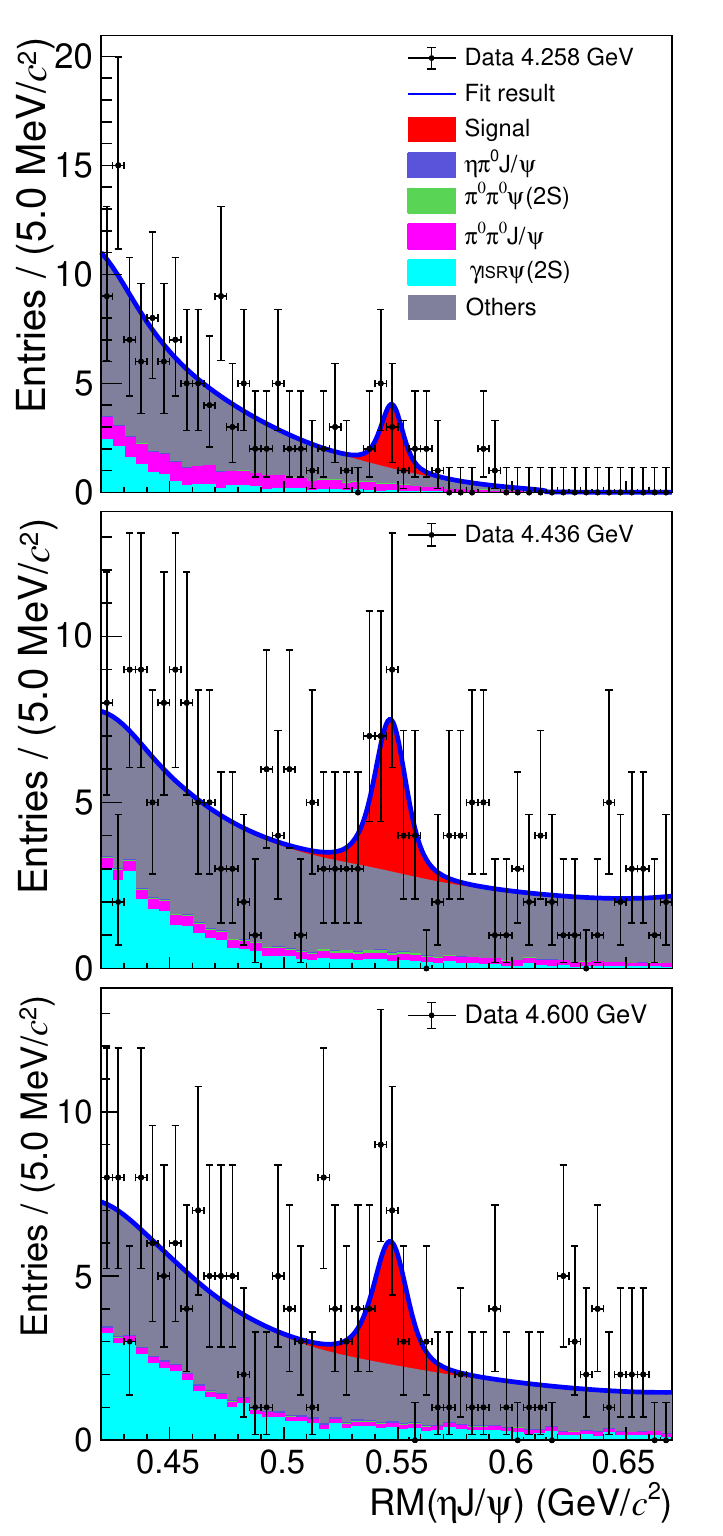}
	\put(41, 31){\bf\large c}
	\put(41, 62){\bf\large b}
	\put(41, 93.5){\bf\large a}
	\end{overpic} 
	\caption{Fits to the $RM(\jpsi\eta)$ distributions at (a) $\sqrt{s} = 4.258$~GeV, (b) $\sqrt{s} = 4.436$~GeV and (c) $\sqrt{s} = 4.600$~GeV. The dots with error bars are the data, the blue curves are the fit results, the red filled histograms are the signal MC samples, the other filled histograms are the normalized MC samples as described in the legenda.
	}
	\label{fig:fitdata}
\end{figure}

\begin{table*}[htbp]
  \centering
	\addtolength{\tabcolsep}{5pt}
  \caption{
  Born cross sections $\sigma^{\rm Born}$ and upper limits at the 90\% CL $\sigma^{\rm UL}$ of the semi-inclusive method for the reaction channel $\elp \elm \to \eta \eta \jpsi$ as a function the CM energy $\sqrt{s}$. The first uncertainty of the Born cross sections $\sigma^{\rm Born}$ is statistical, and the second one systematic. The values needed for the $\sigma^{\rm Born}$ calculation:
  time-integrated luminosity $\mathcal{L}$, event selection efficiency $\epsilon$, signal yield $N_{\rm sig}$, ISR correction factor $(1+\delta)$, and vacuum polarization correction factor $\frac{1}{|1-\Pi|^{2}}$. The statistical significance of $\sigma^{\rm Born}$ is $\mathcal{SS}$. The uncertainties of $N_{\rm sig}$ are only statistical. $N_{\rm UL}$ is the upper limit on the number of events.}
  \begin{tabular}{cr|crc|cc|ccc}
    \hline
    \hline  \\[-1em]
    $\sqrt{s}$ (GeV) &$\mathcal{L}$ (pb$^{-1}$) &$\epsilon$ &$N_{\rm sig}$\;\;\;\;  &$N_{\rm UL}$\;\;\;  &$(1+\delta)$ &$\frac{1}{|1-\Pi|^{2}}$ &$\sigma^{\rm Born}$ (pb) &$\mathcal{SS}$ ($\sigma$) &$\sigma^{\rm UL}$ (pb)  \\  \\[-1em]
    \hline \\[-1em]
    4.226&1100.9\;\;  &0.215 &\;\;$0.62 _{-1.59 }^{+2.09 }$ & 5.2\;\;     &0.674  &1.056 &$ 0.03_{-0.08}^{+0.10}\pm0.01$       &0.3 &0.3   \\ \\[-1em]
    4.236 &530.3\;\;  &0.210 &\;\;$4.96 _{-2.71 }^{+3.26 }$ &11.4\;\;     &0.691  &1.056 &$ 0.51_{-0.28}^{+0.34}\pm0.08$       &2.0 &1.2   \\ \\[-1em]
    4.244 &538.1\;\;  &0.207 &\;\;$7.36 _{-2.95 }^{+3.66 }$ &14.8\;\;     &0.713  &1.056 &$ 0.73_{-0.29}^{+0.37}\pm0.12$       &3.2 &1.5   \\ \\[-1em]
    4.258 &828.4\;\;  &0.206 &\;\;$8.60 _{-3.98 }^{+4.70 }$ &18.3\;\;     &0.739  &1.054 &$ 0.54_{-0.25}^{+0.30}\pm0.09$       &2.4 &1.2   \\ \\[-1em]
    4.267 &531.1\;\;  &0.205 &\;\;$3.14 _{-3.77 }^{+4.60 }$ &12.5\;\;     &0.751  &1.053 &$ 0.31_{-0.37}^{+0.45}\pm0.05$       &0.8 &1.2   \\ \\[-1em]
    4.278 &175.7\;\;  &0.200 &\;\;$2.11 _{-1.79 }^{+2.53 }$ & 7.5\;\;     &0.756  &1.053 &$ 0.63_{-0.54}^{+0.76}\pm0.11$       &1.2 &2.3   \\ \\[-1em]
    4.288 &502.4\;\;  &0.205 &\;\;$9.51 _{-4.91 }^{+5.82 }$ &20.8\;\;     &0.764  &1.053 &$ 0.96_{-0.50}^{+0.59}\pm0.16$       &2.0 &2.1   \\ \\[-1em]
    4.312 &501.0\;\;  &0.204 &\;\;$1.71 _{-4.02 }^{+4.66 }$ &11.1\;\;     &0.771  &1.052 &$ 0.17_{-0.41}^{+0.47}\pm0.03$       &0.4 &1.1   \\ \\[-1em]
    4.337 &505.0\;\;  &0.202 &$14.67_{-6.43 }^{+7.15 }$     &28.1\;\;     &0.777  &1.051 &$ 1.48_{-0.65}^{+0.72}\pm0.25$       &2.5 &2.8   \\ \\[-1em]
    4.358 &543.9\;\;  &0.199 &$12.42_{-6.76 }^{+7.50 }$     &27.2\;\;     &0.796  &1.051 &$ 1.15_{-0.63}^{+0.70}\pm0.19$       &1.9 &2.5   \\ \\[-1em]
    4.377 &522.7\;\;  &0.196 &$15.28_{-7.89 }^{+8.67 }$     &31.4\;\;     &0.822  &1.051 &$ 1.45_{-0.75}^{+0.82}\pm0.24$       &2.0 &3.0   \\ \\[-1em]
    4.396 &507.8\;\;  &0.195 &$11.20_{-6.90 }^{+7.71 }$     &25.3\;\;     &0.829  &1.051 &$ 1.09_{-0.67}^{+0.75}\pm0.18$       &1.7 &2.5   \\ \\[-1em]
    4.416&1090.7\;\;  &0.197 &$31.24_{\;\;-9.96 }^{+10.84}$ &53.0\;\;     &0.806  &1.052 &$ 1.44_{-0.46}^{+0.50}\pm0.25$       &3.4 &2.4   \\ \\[-1em]
    4.436 &569.9\;\;  &0.197 &$19.79_{-7.08 }^{+7.82 }$     &35.9\;\;     &0.856  &1.054 &$ 1.64_{-0.59}^{+0.65}\pm0.28$       &3.1 &3.0   \\ \\[-1em]
    4.467 &111.1\;\;  &0.179 &$-1.14_{-2.05 }^{+2.72 }$     & 5.7\;\;     &0.966  &1.055 &$-0.47_{-0.85}^{+1.13}\pm0.08$\;\;\; &0.5 &2.3   \\ \\[-1em]
    4.527 &112.1\;\;  &0.178 &$\;\;2.36 _{-2.30 }^{+3.07 }$ & 8.6\;\;     &0.934  &1.054 &$ 1.00_{-0.98}^{+1.31}\pm0.17$       &1.0 &3.6   \\ \\[-1em]
    4.600 &586.9\;\;  &0.189 &$17.79_{-6.93 }^{+7.72 }$     &32.9\;\;     &0.839  &1.055 &$ 1.52_{-0.59}^{+0.66}\pm0.26$       &2.8 &2.8   \\ \\[-1em]
    4.612 &103.8\;\;  &0.191 &$\;\;1.65 _{-2.93 }^{+3.64 }$ & 8.7\;\;     &0.839  &1.055 &$ 0.79_{-1.40}^{+1.74}\pm0.13$       &0.5 &4.2   \\ \\[-1em]
    4.628 &521.5\;\;  &0.189 &$10.08_{-7.00 }^{+7.94 }$     &22.4\;\;     &0.878  &1.054 &$ 0.93_{-0.64}^{+0.73}\pm0.16$       &1.5 &2.1   \\ \\[-1em]
    4.641 &552.4\;\;  &0.184 &$16.35_{-7.26 }^{+8.02 }$     &30.5\;\;     &0.937  &1.054 &$ 1.37_{-0.61}^{+0.67}\pm0.24$       &2.4 &2.5   \\ \\[-1em]
    4.661 &529.6\;\;  &0.174 &$13.52_{-6.65 }^{+7.45 }$     &26.5\;\;     &1.018  &1.054 &$ 1.15_{-0.56}^{+0.63}\pm0.21$       &2.2 &2.2   \\ \\[-1em]
    4.682&1669.3\;\;  &0.172 &$10.67_{-10.64}^{+11.39}$     &28.8\;\;     &0.995  &1.054 &$ 0.30_{-0.30}^{+0.32}\pm0.05$       &1.0 &0.8   \\ \\[-1em]
    4.698 &536.5\;\;  &0.176 &$17.67_{-7.25 }^{+8.03 }$     &31.9\;\;     &0.960  &1.055 &$ 1.55_{-0.64}^{+0.70}\pm0.26$       &2.7 &2.8   \\ \\[-1em]
    4.740 &164.3\;\;  &0.183 &$\;\;5.46 _{-3.79 }^{+4.53 }$ &13.6\;\;     &0.888  &1.055 &$ 1.63_{-1.13}^{+1.35}\pm0.27$       &1.5 &4.1   \\ \\[-1em]
    4.750 &367.2\;\;  &0.182 &$20.93_{-7.02 }^{+7.80 }$     &35.7\;\;     &0.880  &1.055 &$ 2.84_{-0.95}^{+1.06}\pm0.48$       &3.4 &4.8   \\ \\[-1em]
    4.780 &512.8\;\;  &0.183 &$16.72_{-6.17 }^{+6.98 }$     &29.9\;\;     &0.868  &1.055 &$ 1.63_{-0.60}^{+0.68}\pm0.28$       &3.1 &2.9   \\ \\[-1em]
    4.842 &527.3\;\;  &0.187 &$11.02_{-6.00 }^{+6.87 }$     &22.8\;\;     &0.855  &1.056 &$ 1.04_{-0.56}^{+0.65}\pm0.18$       &1.9 &2.1   \\ \\[-1em]
    4.918 &208.1\;\;  &0.189 &$\;\;7.75 _{-4.74 }^{+5.52 }$ &17.7\;\;     &0.855  &1.056 &$ 1.83_{-1.12}^{+1.30}\pm0.31$       &1.7 &4.2   \\ \\[-1em]
    4.950 &160.4\;\;  &0.186 &$\;\;8.69 _{-4.93 }^{+5.69 }$ &18.5\;\;     &0.851  &1.056 &$ 2.71_{-1.54}^{+1.78}\pm0.46$       &1.8 &5.8   \\ \\[-1em]
    \hline
    \hline
  \end{tabular}
  \label{tab:results}
\end{table*}

The Born cross section $\sigma^{\rm Born}$ is calculated by
\begin{equation}
	\sigma^{\rm Born} = \frac{N_{\rm sig}}{\mathcal{L}\cdot\epsilon\cdot\br_{\jpsi \to \lp \lm}\cdot(1+\delta)\cdot\frac{1}{|1-\Pi|^{2}}},
	\label{eq:borncs}
\end{equation}
where $N_{\rm sig}$ is the number of signal events, $\mathcal{L}$ the time-integrated luminosity, $\epsilon$ is the event selection efficiency which includes the $\eta$ decay branching fractions,
$\br_{\jpsi \to \lp \lm}$ is the branching fraction of $\jpsi$ decaying into lepton pairs as quoted in the PDG~\cite{ParticleDataGroup:2024cfk},
$(1+\delta)$ is the ISR correction factor, and $\frac{1}{|1-\Pi|^{2}}$ is the vacuum polarization factor taken from Ref.~\cite{WorkingGrouponRadiativeCorrections:2010bjp}.

To obtain the number of signal events $N_{\rm sig}$, an unbinned maximum likelihood fit to the $RM(\jpsi\eta)$ distribution is performed
at each energy point individually. The fit function includes the signal and the background shape.
The signal shape is derived from the $RM(\jpsi\eta)$ distribution of the signal MC samples,
while the background distribution is described by an ARGUS function.
In addition, the shapes of the four channels 
\begin{itemize}
\item {$\elp \elm \to \gamma_{\rm ISR}\psip$ ($\psip \to\piz\piz\jpsi$),}
\item {$\elp \elm \to \piz \piz \jpsi$,} 
\item {$\elp \elm \to \piz \piz \psip$ ($\psip \to\piz\piz\jpsi$), and} 
\item {$\elp \elm \to \eta \piz \jpsi$} 
\end{itemize}
are determined with the corresponding exclusive MC samples. They are added to the fit function with a fixed number of events if their cross sections are known~\cite{BESIII:2017vtc, BESIII:2015cld, BESIII:2015aym}. The results of the fits to the data samples at $\sqrt{s} = 4.258$, $4.436$ and $4.600$~GeV are shown in Fig.~\ref{fig:fitdata},
and the obtained values of signal yields $N_{\rm sig}$ and statistical significances ($\mathcal{SS}$) are given in Table~\ref{tab:results}.
Here, the statistical significance is estimated based on the change of likelihood values from the fits with and without including the signal.

The $\epsilon$ and $(1+\delta)$ are estimated based on signal MC samples, and weighted by a dressed cross section iterative weighting method~\cite{Sun:2020ehv}.
In the iterations, we describe the dressed cross sections 
\begin{equation}
\sigma^{\rm dress} = \sigma^{\rm Born}\frac{1}{|1-\Pi|^{2}}
\end{equation}
by a coherent sum of four Breit-Wigner and a phase-space function.
The four Breit-Wigner functions represent the observed $I^{G}J^{PC} = 0^{-}(1^{--})$ resonances $Y(4230)$, $Y(4360)$, $Y(4415)$, and $Y(4660)$~\cite{ParticleDataGroup:2024cfk}. 
The Breit-Wigner function is defined as

\begin{equation}
{\rm BW} \equiv \frac{M}{\sqrt{s}} \frac{\sqrt{12\pi(\Gamma_{ee}\mathcal{B})\Gamma}}{s - M^{2} + iM\Gamma} \sqrt{\frac{\Phi(\sqrt{s})}{\Phi(M)}},
\end{equation}
where
$\Phi(\sqrt{s})$ is the three-body phase-space factor, $\sqrt{s}$ the CM energy, $(\Gamma_{ee}\mathcal{B})$ is the product of the electronic partial width and the branching fraction to $\eta\eta\jpsi$,
while $M$ and $\Gamma$ are the mass and width of the resonance, respectively.
In the iterations, we fix the $M$ and $\Gamma$ of the corresponding resonance, while letting free the other parameters.

The Born cross sections and the detailed quantities, including the weighted $\epsilon$ and $(1+\delta)$ quantities necessary for the calculation, are listed in Table~\ref{tab:results}. At five CM energies the statistical significance is larger than $3\sigma$, and evidence of the reaction channel could be claimed. At all other values, the significance is around 2$\sigma$, thus the upper limits at the 90\% CL are determined. The determination of the systemtic uncertainties are described in section~\ref{incl:syst}.

\subsection{Systematic uncertainty}
\label{incl:syst}
The systematic uncertainties of the cross section measurements are originating from the time-integrated luminosity, the branching fraction of $\jpsi \to \elp\elm / \mup\mum$, 
the fit procedures, the ISR correction factor, and the detection efficiency.

The systematic uncertainty from the luminosity was studied before by using Bhabha events to be 0.6\%~\cite{BESIII:2015qfd, BESIII:2022xii, BESIII:2022ulv}.
The systematic uncertainties from the subsequent branching fractions are quoted from the PDG~\cite{ParticleDataGroup:2024cfk}.

The systematic uncertainties due to the fit procedures result from the fit range, the signal and the background shapes.
To estimate the influence of the fit range, we enlarge or reduce the fit range by 15~MeV/$c^{2}$.
To estimate the influence of the signal shape, we take into consideration the resolution deviation between data and MC samples by a free Gaussian function. 
To estimate the influence of the background shape, we replace the ARGUS function with a Chebychev function to describe the background distribution.
For the determination of the systematic uncertainties of the fit procedures we group the data sets. 
If we would estimate the uncertainties for each CM energy separately, statistical fluctuations would dominate the results due to the very limited signal yield. The differences vary from (8.4 - 9.3)\%, which are assigned as the corresponding systematic uncertainties due to the fit procedures.

To estimate the systematic uncertainty from the ISR correction factor, we compare the dressed cross sections between the last two ISR correction iterations.
The difference in the dressed cross sections is low enough to be neglected.
On the other hand, in the ISR correction iterations, we replace the description of the default dressed cross section line shape
as the various coherent sum of phase-space and several (less than four) Breit-Wigner functions. Various combinations of the resonances 
$Y(4230)$, $Y(4360)$, $Y(4415)$, $Y(4660)$, and phase-space are taken into consideration. 
The maximum difference to the default line shape is (0.6 - 7.7)\%, which is assigned as the systematic uncertainty due to the ISR correction factor.

The systematic uncertainty for the detection efficiency includes the systematic uncertainties
in tracking efficiency (1.0\% per track) and photon reconstruction (1.0\% per photon)~\cite{BESIII:2015wyx}.
In addition, the uncertainty due to the kinematic fit is estimated by the comparison of using and not using the helix parameter corrections of the simulated charged
tracks to match the resolution found in data.
The difference (0.9\%) is assigned as the systematic uncertainty due to the kinematic fit.
The systematic uncertainty due to the requirement on the number of the MUC hits
has been estimated as 3.0\% with a control sample of $\elp \elm \to \piz \piz \jpsi$,
by comparing the efficiency of the MUC requirement between data and MC samples~\cite{BESIII:2015cld}.

The systematic uncertainty due to the MC model is assigned as the maximum difference of the efficiencies between various assumptions of 
intermediate states of the reaction channel $\elp\elm\to\eta\eta\jpsi$. Therefore, the potential intermediate states $\psi(2S)$, $f_{2}(1270)$, $f_{0}(1500)$, $f_{2}^{\prime}(1525)$, 
and $f_{4}(2050)$ are taken into consideration, based on the simulation of the channels
$\elp\elm\to\eta\psi(2S)$ ($\psi(2S)\to\jpsi\eta$), $\elp\elm\to f_{0}(1500)\jpsi$ ($f_{0}(1500)\to\eta\eta$), 
$\elp\elm\to f_{2}(1270)\jpsi$ ($f_{2}(1270)\to\eta\eta$), and $\elp\elm\to f_{2}^{\prime}(1525)\jpsi$ ($f_{2}^{\prime}(1525)\to\eta\eta$), 
where $\jpsi$ decays into a lepton pair. The maximum difference of the efficiencies, 13.4\%, is assigned as the uncertainty due to the MC model.

\begin{table}[htbp]
  \centering\addtolength{\tabcolsep}{5pt}
  \caption{Relative systematic uncertainties in the measurement of the cross section $\elp\elm\to\eta\eta\jpsi$ for all data samples.}
  \begin{tabular}{cc}
    \hline
    \hline
    Source &Systematic uncertainty (\%) \\
    \hline
    Luminosity					&0.6 \\
    $\br_{\jpsi \to \lp \lm}$			&0.4 \\
    Fit procedures     				&8.4-9.3 \\
    $(1+\delta)$       				&0.6-7.7 \\
    Tracking					&2.0 \\
    Photon reconstruction   			&2.0 \\
    Kinematic fit      				&0.9 \\
    $N_{\rm HitsMuc}$  				&3.0 \\
    MC model           				&13.4 \\
    \hline
    Sum                				&16.4-17.6 \\
    \hline
    \hline
  \end{tabular}
  \label{tab:syserr}
\end{table}

Assuming that all the sources are independent, the total systematic uncertainty in the $\elp\elm\to\eta\eta\jpsi$ cross section measurement is determined to be
the quadratic sum of the systematic uncertainties.
Table~\ref{tab:syserr} gives the list of all the relative systematic uncertainties and their sum.

\subsection{Upper limit determination}

According to the Bayesian method~\cite{Feldman:1997qc}, we scan the number of signal events to obtain an area-normalized likelihood distribution, 
and smear the systematic uncertainties into the likelihood distribution. The upper limit $N_{\rm UL}$ of the number of signal events at the 90\% CL is determined by 

\begin{equation}
\frac{\int^{N_{\rm UL}}_{0}F(x){\rm d}x}{\int^{\infty}_{0}F(x){\rm d}x} = 0.90,
\end{equation}
where $F(x)$ is the probability density function of the likelihood distribution smeared with the systematic uncertainties, and $x$ is the scanned number of signal events. 
The upper limit $\sigma^{\rm UL}$ of the Born cross section at 90\% CL  is calculated by replacing $N_{\rm sig}$ of equation~\ref{eq:borncs} as $N_{\rm UL}$.

\begin{figure}[htbp]
	\centering
	\includegraphics[width = 0.43\textwidth]{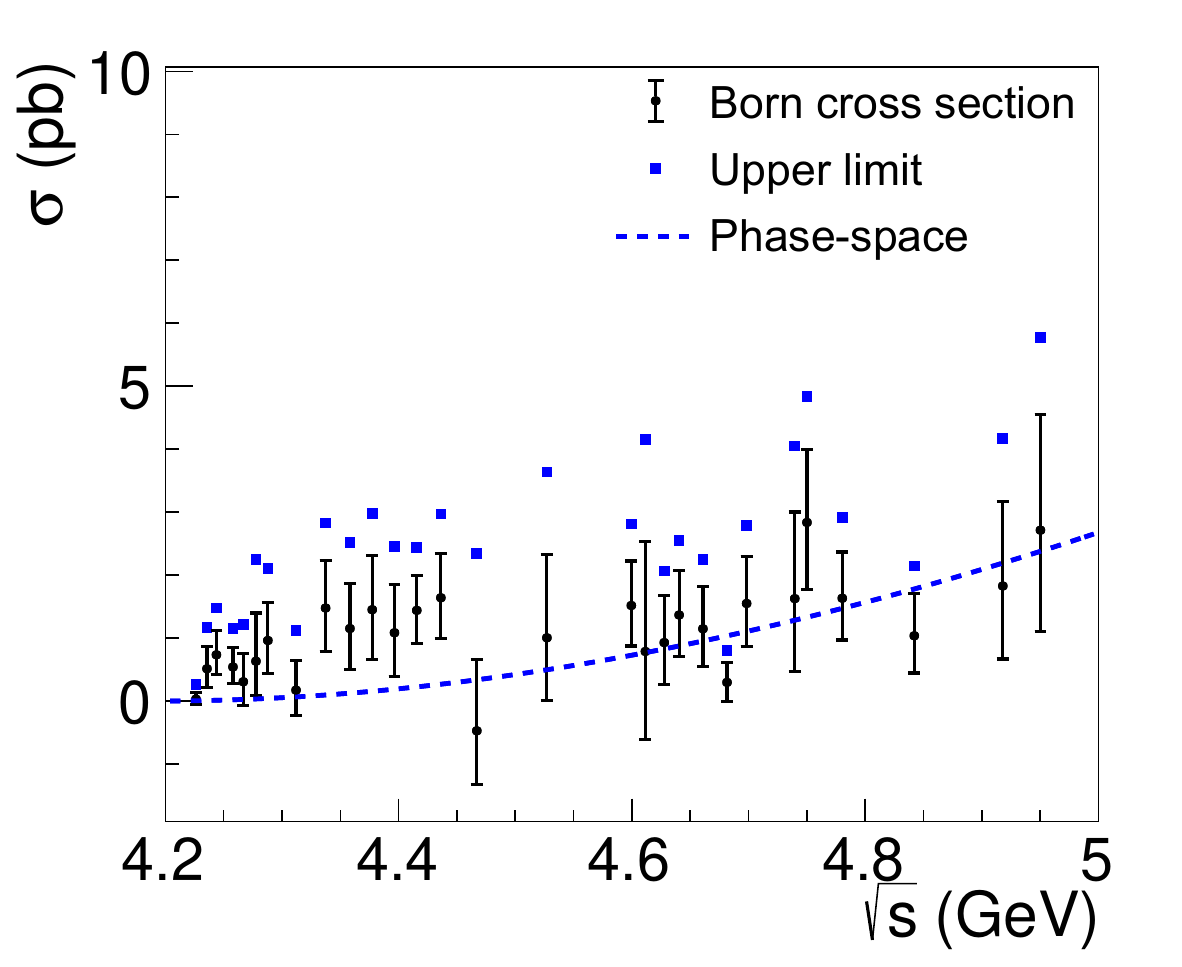}
	\caption{Born cross sections (black) and upper limits (blue dots) at the 90\% CL of the reaction channel $\elp\elm\to\eta\eta\jpsi$ as a function of the CM energy.
		The error bars include the statistical and systematic uncertainties. The blue dashed curve is the calculated three-body phase space.} 
	\label{fig:csfit}
\end{figure}

The detailed quantities of the Born cross sections calculated with equation~(\ref{eq:borncs}) and the upper limits of the cross sections at the 90\% CL are given in Table~\ref{tab:results}.
Figure~\ref{fig:csfit} shows the Born cross sections and their upper limits at the 90\% CL, as well as the corresponding three-body phase-space line shape.
The three-body phase-space line shape describes the overall trend of the Born cross sections as a function of the CM energy quite well. No clear structures are visible due to intermediate resonances.

\subsection{Summary --- Semi-inclusive method}
In this section, we search for the reaction channel $\elp\elm\to\eta\eta\jpsi$ at $\sqrt{s}$ from 4.226 to 4.950~GeV using the data samples corresponding to a time-integrated luminosity of 15.1 fb$^{-1}$ taken at twenty-nine CM energy points.
As shown in Table~\ref{tab:results}, the statistical significance at all CM energies is less than 3.5$\sigma$. This reflects that in most of the data samples no clear $\eta$ signals are observed. Only at five CM energies an evidence could be claimed.
The Born cross sections and the upper limits at the 90\% CL are determined and shown in Fig.~\ref{fig:csfit}. 
No clear structure is observed. The trend of the distribution can be described by a pure phase-space model, but additional contributions by intermediate resonances are needed.

\section{Combination of the results}

\begin{figure}[htbp]
  \centering
  \includegraphics[width=0.43\textwidth]{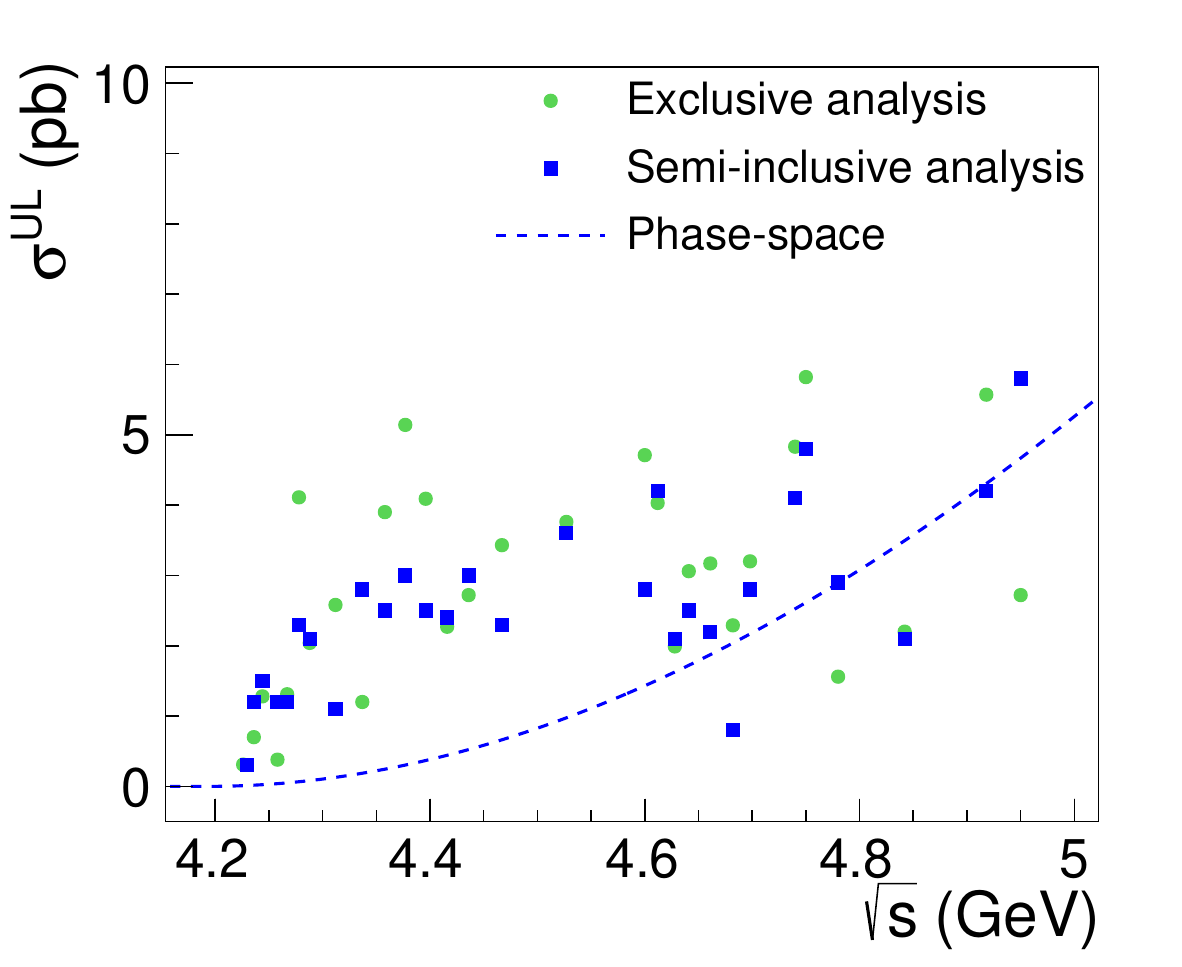}
  \caption{Comparison of the upper limits at the 90\% CL of the Born cross sections of the two analysis methods as a function of the CM energy. The green dots are the upper limits obtained from the exclusive method, 
  while the blue boxes are these from the semi-inclusive method.
  The blue dashed line represents the three-body phase space.
  }
  \label{fig:combined}
\end{figure}

For the combination of the upper limits of both analyses for each energy point, it would be necessary to combine the likelihood distributions of both methods in order to get the combined upper limits at 90\% CL. Since the exclusive method uses a frequentist approach, there is no likelihood distribution accessible, which could be combined with the upper limits from the semi-inclusive approach. Therefore, we give both results next to each other. 
Table~\ref{tab:combined} lists both results at each CM energy. Also both are plotted in Fig.~\ref{fig:combined} together with the pure phase-space line shape. No clear structure can be seen.

\begin{table}[htbp]
  \caption{Upper limits on the Born cross sections at the 90\% CL of the reaction channel $\elp \elm \to \eta\eta\jpsi$ for all 29 CM energies $\sqrt{s}$: the upper limits of the exclusive method $\sigma^{\rm UL}_{\rm excl}$ and the upper limits of the semi-inclusive method $\sigma^{\rm UL}_{\rm semi-incl}$. }
  \label{tab:combined}
  \centering\addtolength{\tabcolsep}{3.0pt}
  \begin{tabular}{c|cc}
    \hline\hline
    $\sqrt{s}~(\si{GeV})$ & $\sigma^{\rm UL}_{\rm excl}~(\si{\pb})$ & $\sigma^{\rm UL}_{\rm semi-incl}~(\si{\pb})$ \\
    \hline
\input{FinalResults}
    \\\hline\hline
  \end{tabular}
\end{table}

\section{Conclusion}

Two analyses with independent methods are performed to search for the reaction channel $\elp\elm\to\eta\eta\jpsi$ at CM energies from 4.226 to 4.950~GeV
using data samples corresponding to a time-integrated luminosity of 15.1 fb$^{-1}$ taken at twenty-nine energy values. With the first method we reconstruct all final state particles exclusively by reconstructing both $\eta$ mesons via $\twogg$ or $\threepi$.  Both observed and Born cross sections are provided.
Evidence of the reaction channel with significance larger than 3$\sigma$ is found at six CM energies, and an observation with significance larger than 5$\sigma$ is made at 4.682 (5.7$\sigma$) and 4.750~GeV (8.9$\sigma$). Since the background situation is still not completely solved, these significances should be taken with caution.
With the second method we reconstruct only one $\eta$ in the decay $\twogg$ and the second $\eta$ is inferred by the missing mass technique. We determine the Born cross section. At five CM energies the significance is larger than 3$\sigma$. 

Since the ISR correction is determined in a completely different way, the Born cross section measurements are not combined. Therefore, for the exclusive method the observed cross sections are provided in addition. The comparison of the results of both methods shows that only at 4.750~GeV both analyses agree to claim an observation (exclusive analysis 8.9$\sigma$) and evidence (semi-inclusive analysis 3.4$\sigma$). At all other CM energies we do not obtain results with significance above 3$\sigma$ from both analyses. Therefore, no observation or evidence is claimed and the upper limits at 90\% confidence level (CL) of the Born cross sections are the final results.

The  upper limits obtained using both methods show a strong deviation from pure phase-space around 4.4~GeV. However, no clear structures are visible.

Since no significant $\elp\elm\to\eta\eta\jpsi$ signal is observed in either of the two analysis methods, and the number of events is limited, no resonant substructure in the final state can be investigated.  The upper limits on 
$\sigma(\elp\elm\to\eta\eta\jpsi)$ listed in Table~\ref{tab:combined} and shown in 
Fig.~\ref{fig:combined} can therefore also be considered upper limits on the product
$\sigma (\elp\elm\to\eta Z'_{\rm c}) \cdot \mathcal{B}(Z'_{\rm c} \to \eta J/\psi)$ for the production and decay 
of an isoscalar exotic state $Z'_{\rm c}$.  To draw clear conclusions, data samples with larger statistics are needed.

\section{Acknowledgements}

\input{acknowledgement.tex}

\FloatBarrier

\bibliography{newreference}
\end{document}

%% file: Authorlist.tex
\author{
M.~Ablikim$^{1}$\BESIIIorcid{0000-0002-3935-619X},
M.~N.~Achasov$^{4,b}$\BESIIIorcid{0000-0002-9400-8622},
P.~Adlarson$^{77}$\BESIIIorcid{0000-0001-6280-3851},
X.~C.~Ai$^{82}$\BESIIIorcid{0000-0003-3856-2415},
R.~Aliberti$^{36}$\BESIIIorcid{0000-0003-3500-4012},
A.~Amoroso$^{76A,76C}$\BESIIIorcid{0000-0002-3095-8610},
Q.~An$^{73,59,\dagger}$,
Y.~Bai$^{58}$\BESIIIorcid{0000-0001-6593-5665},
O.~Bakina$^{37}$\BESIIIorcid{0009-0005-0719-7461},
Y.~Ban$^{47,g}$\BESIIIorcid{0000-0002-1912-0374}
% ...
\\
\vspace{0.2cm}
(BESIII Collaboration)\\
\vspace{0.2cm} {\it
$^{1}$ Institute of High Energy Physics, Beijing 100049, People's Republic of China\\
% ...
$^{4}$ Budker Institute of Nuclear Physics SB RAS (BINP), Novosibirsk 630090, Russia\\
% ...
$^{36}$ Johannes Gutenberg University of Mainz, Johann-Joachim-Becher-Weg 45, D-55099 Mainz, Germany\\
$^{37}$ Joint Institute for Nuclear Research, 141980 Dubna, Moscow region, Russia\\
% ...
$^{47}$ Peking University, Beijing 100871, People's Republic of China\\
% ...
$^{58}$ Southeast University, Nanjing 211100, People's Republic of China\\
$^{59}$ State Key Laboratory of Particle Detection and Electronics, Beijing 100049, Hefei 230026, People's Republic of China\\
% ...
$^{73}$ University of Science and Technology of China, Hefei 230026, People's Republic of China\\
% ...
$^{76}$ University of Turin and INFN, (A)University of Turin, I-10125, Turin, Italy; (B)University of Eastern Piedmont, I-15121, Alessandria, Italy; (C)INFN, I-10125, Turin, Italy\\
$^{77}$ Uppsala University, Box 516, SE-75120 Uppsala, Sweden\\
% ...
$^{82}$ Zhengzhou University, Zhengzhou 450001, People's Republic of China\\
\vspace{0.2cm}
$^{\dagger}$ Deceased\\
% ...
$^{b}$ Also at the Novosibirsk State University, Novosibirsk, 630090, Russia\\
% ...
$^{g}$ Also at State Key Laboratory of Nuclear Physics and Technology, Peking University, Beijing 100871, People's Republic of China\\
% ...
}
}

%% file: crosssection.tex
4.226 & 1100.9 && 0.316(1) & 0.143(1) & 0.057(0) &  0 & 0.21     &  -0.00 & 0.00 & $-0.03^{+0.13}_{-0.00} \pm 0.00$ & $-0.04^{+0.20}_{-0.01} \pm 0.00$ & 0   & 0.330 & 0.672  &0.3\\ \\[-1em]
4.236 &  530.3 && 0.310(1) & 0.140(1) & 0.055(0) &  0 & 0.11     &  -0.00 & 0.00 & $-0.03^{+0.29}_{-0.00} \pm 0.00$ & $-0.04^{+0.43}_{-0.01} \pm 0.00$ & 0   & 0.153 & 0.672  &0.7\\ \\[-1em]
4.244 &  538.1 && 0.307(1) & 0.139(1) & 0.055(0) &  1 & 0.18     &  0.01 & 0.00  & $0.20^{+0.57}_{-0.20} \pm 0.01$  & $ 0.30^{+0.85}_{-0.30} \pm 0.01$ & 1.0 & 0.155 & 0.672  &1.3\\ \\[-1em]
4.258 &  828.4 && 0.304(1) & 0.139(1) & 0.057(0) &  0 & 0.41     &  -0.00 & 0.00 & $-0.07^{+0.19}_{-0.01} \pm 0.00$ & $-0.10^{+0.28}_{-0.01} \pm 0.01$ & 0   & 0.241 & 0.670  &0.4\\ \\[-1em]
4.267 &  531.1 && 0.301(1) & 0.136(1) & 0.055(0) &  1 & 0.22     &  0.01 & 0.01  & $0.20^{+0.59}_{-0.21} \pm 0.01$  & $ 0.30^{+0.88}_{-0.32} \pm 0.01$ & 0.8 & 0.152 & 0.672  &1.3\\ \\[-1em]
4.278 &  175.7 && 0.298(1) & 0.134(1) & 0.054(0) &  1 & 0.12     &  0.03 & 0.01  & $0.69^{+1.80}_{-0.65} \pm 0.03$  & $ 1.02^{+2.67}_{-0.96} \pm 0.04$ & 1.2 & 0.051 & 0.672  &4.1\\ \\[-1em]
4.288 &  502.4 && 0.297(1) & 0.128(1) & 0.049(0) &  2 & 0.39     &  0.02 & 0.02  & $0.45^{+0.73}_{-0.36} \pm 0.03$  & $ 0.67^{+1.09}_{-0.53} \pm 0.04$ & 1.6 & 0.141 & 0.672  &2.0\\ \\[-1em]
4.312 &  501.0 && 0.290(1) & 0.127(1) & 0.050(0) &  3 & 0.74     &  0.02 & 0.13  & $0.64^{+0.83}_{-0.46} \pm 0.13$  & $ 0.96^{+1.24}_{-0.69} \pm 0.20$ & 1.8 & 0.137 & 0.669  &2.6\\ \\[-1em]
4.337 &  505.0 && 0.288(1) & 0.128(1) & 0.051(0) &  1 & 0.80     &  0.00 & 0.06  & $0.06^{+0.65}_{-0.23} \pm 0.06$  & $ 0.08^{+0.97}_{-0.35} \pm 0.08$ & 0   & 0.138 & 0.669  &1.2\\ \\[-1em]
4.358 &  543.9 && 0.290(1) & 0.133(1) & 0.056(0) &  6 & 0.89     &  0.05 & 0.07  & $1.32^{+0.92}_{-0.61} \pm 0.08$  & $ 1.97^{+1.38}_{-0.92} \pm 0.13$ & 3.4 & 0.151 & 0.668  &3.9\\ \\[-1em]
4.377 &  522.7 && 0.284(1) & 0.127(1) & 0.051(0) &  8 & 1.19     &  0.07 & 0.16  & $1.88^{+1.09}_{-0.76} \pm 0.18$  & $ 2.81^{+1.63}_{-1.14} \pm 0.26$ & 4.0 & 0.143 & 0.668  &5.1\\ \\[-1em]
4.396 &  507.8 && 0.283(1) & 0.125(1) & 0.051(0) &  6 & 1.43     &  0.05 & 0.39  & $1.30^{+1.02}_{-0.68} \pm 0.39$  & $ 1.95^{+1.53}_{-1.02} \pm 0.58$ & 2.7 & 0.136 & 0.668  &4.1\\ \\[-1em]
4.416 & 1090.7 && 0.281(1) & 0.129(1) & 0.057(0) &  8 & 2.14     &  0.03 & 0.12  & $0.77^{+0.52}_{-0.37} \pm 0.13$  & $ 1.16^{+0.78}_{-0.55} \pm 0.19$ & 2.9 & 0.291 & 0.670  &2.3\\ \\[-1em]
4.436 &  569.9 && 0.281(1) & 0.127(1) & 0.052(0) &  4 & 1.11     &  0.03 & 0.08  & $0.74^{+0.81}_{-0.49} \pm 0.09$  & $ 1.10^{+1.20}_{-0.73} \pm 0.13$ & 1.9 & 0.150 & 0.670  &2.7\\ \\[-1em]
4.467 &  111.1 && 0.278(1) & 0.129(1) & 0.057(0) &  0 & 0.24     &  -0.01 & 0.23 & $-0.31^{+1.51}_{-0.01} \pm 0.23$ & $-0.46^{+2.24}_{-0.01} \pm 0.35$ & 0   & 0.029 & 0.671  &3.4\\ \\[-1em]
4.527 &  112.1 && 0.277(1) & 0.129(1) & 0.058(0) &  0 & 0.05     &  -0.00 & 0.11 & $-0.07^{+1.49}_{-0.00} \pm 0.11$ & $-0.10^{+2.23}_{-0.00} \pm 0.16$ & 0   & 0.031 & 0.671  &3.8\\ \\[-1em]
4.600 &  586.9 && 0.276(1) & 0.131(1) & 0.061(0) &  8 & 0.87     &  0.07 & 0.06  & $1.77^{+0.98}_{-0.69} \pm 0.09$  & $ 2.63^{+1.46}_{-1.02} \pm 0.13$ & 4.5 & 0.154 & 0.671  &4.7\\ \\[-1em]
4.612 &  103.8 && 0.275(1) & 0.129(1) & 0.056(0) &  0 & 0.08     &  -0.00 & 0.12 & $-0.11^{+1.62}_{-0.00} \pm 0.12$ & $-0.16^{+2.42}_{-0.00} \pm 0.19$ & 0   & 0.027 & 0.671  &4.0\\ \\[-1em]
4.628 &  521.5 && 0.275(1) & 0.130(1) & 0.058(0) &  2 & 0.55     &  0.02 & 0.04  & $0.41^{+0.74}_{-0.36} \pm 0.04$  & $ 0.60^{+1.10}_{-0.54} \pm 0.07$ & 1.2 & 0.139 & 0.671  &2.0\\ \\[-1em]
4.641 &  552.4 && 0.275(1) & 0.130(1) & 0.058(0) &  4 & 0.52     &  0.04 & 0.05  & $0.92^{+0.84}_{-0.51} \pm 0.06$  & $ 1.37^{+1.25}_{-0.76} \pm 0.09$ & 2.9 & 0.143 & 0.671  &3.1\\ \\[-1em]
4.661 &  529.6 && 0.275(1) & 0.130(1) & 0.059(0) &  4 & 0.55     &  0.04 & 0.03  & $0.95^{+0.87}_{-0.53} \pm 0.05$  & $ 1.42^{+1.30}_{-0.79} \pm 0.07$ & 2.8 & 0.139 & 0.671  &3.2\\ \\[-1em]
4.682 & \;\;1669.3 && 0.274(1) & 0.130(1) & 0.059(0) & 12 & 1.16 &  0.03 & 0.01  & $0.95^{+0.40}_{-0.30} \pm 0.04$  & $ 1.42^{+0.60}_{-0.45} \pm 0.06$ & 5.7 & 0.413 & 0.671  &2.3\\ \\[-1em]
4.698 &  536.5 && 0.274(1) & 0.131(1) & 0.060(0) &  4 & 0.40     &  0.04 & 0.03  & $0.98^{+0.86}_{-0.52} \pm 0.05$  & $ 1.46^{+1.28}_{-0.78} \pm 0.07$ & 3.2 & 0.144 & 0.671  &3.2\\ \\[-1em]
4.740 &  164.3 && 0.273(1) & 0.132(1) & 0.061(0) &  1 & 0.01     &  0.03 & 0.04  & $0.88^{+2.05}_{-0.74} \pm 0.05$  & $ 1.31^{+3.05}_{-1.10} \pm 0.07$ & 2.2 & 0.043 & 0.671  &4.8\\ \\[-1em]
4.750 &  367.2 && 0.267(1) & 0.131(1) & 0.062(0) &  5 & 0.00     &  0.07 & 0.02  & $2.02^{+1.37}_{-0.87} \pm 0.08$  & $ 3.01^{+2.04}_{-1.30} \pm 0.12$ & 8.9 & 0.091 & 0.671  &5.8\\ \\[-1em]
4.780 &  512.8 && 0.268(1) & 0.131(1) & 0.061(0) &  1 & 0.04     &  0.01 & 0.02  & $0.28^{+0.67}_{-0.24} \pm 0.02$  & $ 0.42^{+0.99}_{-0.36} \pm 0.04$ & 1.8 & 0.127 & 0.671  &1.6\\ \\[-1em]
4.842 &  527.3 && 0.269(1) & 0.134(1) & 0.064(1) &  2 & 0.02     &  0.02 & 0.02  & $0.55^{+0.73}_{-0.36} \pm 0.03$  & $ 0.82^{+1.09}_{-0.54} \pm 0.04$ & 3.6 & 0.136 & 0.672  &2.2\\ \\[-1em]
4.918 &  208.1 && 0.268(1) & 0.136(1) & 0.066(1) &  2 & 0.01     &  0.05 & 0.02  & $1.40^{+1.86}_{-0.91} \pm 0.06$  & $ 2.09^{+2.77}_{-1.36} \pm 0.08$ & 4.0 & 0.054 & 0.672  &5.6\\ \\[-1em]
4.950 &  160.4 && 0.268(1) & 0.134(1) & 0.067(1) &  0 & 0.00     &  -0.00 & 0.02 & $-0.00^{+1.05}_{-0.00} \pm 0.02$ & $-0.00^{+1.57}_{-0.00} \pm 0.03$ & 0   & 0.040 & 0.672  &2.7

%% file: FinalResults.tex
4.226 &0.3 &0.3  \\
4.236 &0.7 &1.2  \\
4.244 &1.3 &1.5  \\
4.258 &0.4 &1.2  \\
4.267 &1.3 &1.2  \\
4.278 &4.1 &2.3  \\
4.288 &2.0 &2.1  \\
4.312 &2.6 &1.1  \\
4.337 &1.2 &2.8  \\
4.358 &3.9 &2.5  \\
4.377 &5.1 &3.0  \\
4.396 &4.1 &2.5  \\
4.416 &2.3 &2.4  \\
4.436 &2.7 &3.0  \\
4.467 &3.4 &2.3  \\
4.527 &3.8 &3.6  \\
4.600 &4.7 &2.8  \\
4.612 &4.0 &4.2  \\
4.628 &2.0 &2.1  \\
4.641 &3.1 &2.5  \\
4.661 &3.2 &2.2  \\
4.682 &2.3 &0.8  \\
4.698 &3.2 &2.8  \\
4.740 &4.8 &4.1  \\
4.750 &5.8 &4.8  \\
4.780 &1.6 &2.9  \\
4.842 &2.2 &2.1  \\
4.918 &5.6 &4.2  \\
4.950 &2.7 &5.8

%% file: acknowledgement.tex

The BESIII Collaboration thanks the staff of BEPCII (https://cstr.cn/31109.02.BEPC) and the IHEP computing center for their strong support. This work is supported in part by National Key R\&D Program of China under Contracts Nos. 2025YFA1613900, 2023YFA1606000, 2023YFA1606704; National Natural Science Foundation of China (NSFC) under Contracts Nos. 11635010, 11935015, 11935016, 11935018, 12025502, 12035009, 12035013, 12061131003, 12192260, 12192261, 12192262, 12192263, 12192264, 12192265, 12221005, 12225509, 12235017, 12361141819; the Chinese Academy of Sciences (CAS) Large-Scale Scientific Facility Program; CAS under Contract No. YSBR-101; 100 Talents Program of CAS; The Institute of Nuclear and Particle Physics (INPAC) and Shanghai Key Laboratory for Particle Physics and Cosmology; German Research Foundation DFG under Contract No. FOR5327; Istituto Nazionale di Fisica Nucleare, Italy; Knut and Alice Wallenberg Foundation under Contracts Nos. 2021.0174, 2021.0299; Ministry of Development of Turkey under Contract No. DPT2006K-120470; National Research Foundation of Korea under Contract No. NRF-2022R1A2C1092335; National Science and Technology fund of Mongolia; National Science Research and Innovation Fund (NSRF) via the Program Management Unit for Human Resources \& Institutional Development, Research and Innovation of Thailand under Contract No. B50G670107; Polish National Science Centre under Contract No. 2024/53/B/ST2/00975; Swedish Research Council under Contract No. 2019.04595; U. S. Department of Energy under Contract No. DE-FG02-05ER41374
